\documentclass[10pt]{iopart}

\usepackage[utf8]{inputenc} %
\usepackage[T1]{fontenc}    %
\usepackage{hyperref}       %
\usepackage{url}            %
\usepackage{booktabs}       %
\usepackage{amsfonts}       %
\usepackage{nicefrac}       %
\usepackage{microtype}      %
\usepackage[table, x11names, dvipsnames]{xcolor}         %

\usepackage{siunitx} %
\usepackage{fontawesome5}  %

\usepackage{amsmath} %
\usepackage{mathtools}

\usepackage{url} %
\usepackage{float}

\usepackage{caption}
\usepackage{subcaption}

\usepackage{listings} %

\usepackage{booktabs}

\definecolor{amaranth}{rgb}{0.9, 0.17, 0.31}
\colorlet{green}{green!20}
\colorlet{yellow}{yellow!60}
\colorlet{red}{red!30}

\usepackage{setspace}
\usepackage[linesnumbered,ruled,vlined]{algorithm2e}

\SetKwComment{Comment}{\color{green!80!black!190}// }{}

\SetKwProg{Function}{function}{}{}
\SetKw{Continue}{continue}

\DontPrintSemicolon
\SetAlFnt{\small}
\SetAlgorithmName{Pseudocode}{algorithmautorefname}

\usepackage{tikz}
\usetikzlibrary{decorations.pathreplacing,calc}

\usepackage{enumitem}

\usepackage{siunitx} %

\usepackage{diagbox} %

\usepackage{multirow}  

\usepackage{longtable} %

\usepackage{array}
\newcolumntype{?}[1]{!{\vrule width #1}}
\def\showcomments{} %

\ifx\showcomments\undefined
\newcommand{\CV}[1]{}
\newcommand{\YP}[1]{}
\else
\newcommand{\CV}[1]{\textcolor{red}{[CV: #1]}}
\newcommand{\YP}[1]{\textcolor{purple}{[YP: #1]}}
\fi

\usepackage[absolute,overlay]{textpos}
\usepackage{times}

\usepackage{scalerel}
\newcommand\vfrac[2]{\ThisStyle{%
  \setbox0=\hbox{$\SavedStyle#1#2$}%
  \setbox2=\hbox{$\SavedStyle X$}%
  \ifdim\ht0>\ht2\setlength{\ht0}{\ht2}\fi%
  #1\mathord{\stretchto{\raisebox{2.3\LMpt}{$\SavedStyle/$}}{\ht0}}#2}}

\usepackage{cite}

\begin{document}

\ioptwocolmod
\twocolumn[{
\begin{@twocolumnfalse}
\title[Robust Confinement State Classification with Uncertainty Quantification through Ensembled Data-Driven Methods]{Robust Confinement State Classification with Uncertainty Quantification through Ensembled Data-Driven Methods}
\vspace{-2pt}
\author{Yoeri Poels$^{\text{\textdagger},1,2}$, Cristina Venturini$^{\text{\textdagger},1}$, Alessandro Pau$^{1}$, Olivier Sauter$^{1}$, Vlado Menkovski$^{2}$, the TCV team$^3$ and the WPTE team$^{4}$}
\address{$^1$École Polytechnique Fédérale de Lausanne, Swiss Plasma Center, CH-1015 Lausanne, Switzerland}
\address{$^2$Eindhoven University of Technology, Mathematics and Computer Science, NL-5600MB Eindhoven, The Netherlands}
\address{$^3$See author list of B. P. Duval \textit{et al.} 2024 \textit{Nucl. Fusion} \textbf{64} 112023}
\address{$^4$See author list of E. Joffrin \textit{et al.} 2024 \textit{Nucl. Fusion} \textbf{64} 112019}
\address{$^{\text{\textdagger}}$Equal contribution, sorted alphabetically.}
\vspace{-2.6pt}
\ead{yoeri.poels@epfl.ch, cristina.venturini@epfl.ch}
\vspace{1.4pt}
\begin{indented}
\item[]February 2025
\end{indented}
\textbf{Abstract}\\
Maximizing fusion performance in tokamaks relies on high energy confinement, often achieved through distinct operating regimes. The automated labeling of these confinement states is crucial to enable large-scale analyses or for real-time control applications. While this task becomes difficult to automate near state transitions or in marginal scenarios, much success has been achieved with data-driven models. However, these methods generally provide predictions as point estimates, and cannot adequately deal with missing and/or broken input signals. To enable wide-range applicability, we develop methods for confinement state classification with \textit{uncertainty quantification} and \textit{model robustness}. We focus on off-line analysis for TCV discharges, distinguishing L-mode, H-mode, and an in-between dithering phase (D). We propose ensembling data-driven methods on two axes: model formulations and feature sets. The former considers a dynamic formulation based on a recurrent Fourier Neural Operator-architecture and a static formulation based on gradient-boosted decision trees. These models are trained using multiple feature groupings categorized by diagnostic system or physical quantity. A dataset of 302 TCV discharges is fully labeled, and we release it publicly to encourage the community to build upon this work. We evaluate our method quantitatively using Cohen's kappa coefficient for predictive performance and the Expected Calibration Error for the uncertainty calibration. Furthermore, we discuss performance using a variety of common and alternative scenarios, the performance of individual components, out-of-distribution performance, cases of broken or missing signals, and evaluate conditionally-averaged behavior around different state transitions. Overall, the proposed method can distinguish L, D and H-mode with high performance, can cope with missing or broken signals, and provides meaningful uncertainty estimates. 

\vspace{.13cm}
\hrule
\vspace{.13cm}
\end{@twocolumnfalse}
}\vspace{-2.9cm}]

\renewcommand*{\thefootnote}{\arabic{footnote}}

\maketitle

\section{Introduction}\label{sec:intro}
In magnetic confinement fusion, the energy confinement time of the plasma is one of the key parameters for maximizing fusion performance. This quantity is known to scale with various plasma parameters, for example the plasma shape, the particle density or the strength of the magnetic field, among others~\cite{itpascaling1990}. However, distinct operating regimes have been discovered that provide better-than-expected scaling, which fall under the umbrella term of high-confinement mode (H-mode) regimes~\cite{hmode1982}. Operating in these high-performance regimes is crucial to maximize the performance of current-day and future devices~\cite{hmodeiter2008}.

To accelerate large-scale analysis of confinement states, or for real time control-scenarios, we need automatic confinement state detection algorithms. This task becomes difficult near state transitions or in marginal scenarios, however, much success has been achieved with data-driven models. Past works have developed methods for full-discharge confinement state identification on Alcator C-Mod~\cite{mathews2019}, COMPASS~\cite{zorek2022}, DIII-D~\cite{orozco2022,gill2024}, EAST~\cite{yang2024}, HL-2A~\cite{he2024}, KSTAR~\cite{shin2018,shin2020}, JET~\cite{meakins2010,gonzalez2012}, and TCV~\cite{matoslhd2020,matoslhd2021}. 

Still, these methods generally do not consider two key aspects. For one, predictions are generally provided as point estimates, giving no information about the associated prediction uncertainty. This additional dimension is critical to identify when model predictions can be trusted, for example in control scenarios or to ensure high-quality analyses. Additionally, the ability to deal with missing and/or broken input signals is generally not addressed. To enable wide-range applicability, models must be robust to these failure modes. Notably, some related works do incorporate a notion of uncertainty~\cite{gonzalez2012,verdoolaege2012,verdoolaege2012ppcf}, however, not in the full discharge setting or with expressive models such as neural networks (NNs). 

To incorporate the notions of \textit{uncertainty quantification} and \textit{model robustness}, we propose the use of ensembled data-driven methods. We combine methods on two axes: different types of models, and different sets of input signals. The former allows us to incorporate different inductive biases, i.e. varying the assumptions made by the algorithm. As a consequence we expect to reduce failure modes connected to model properties~\cite{strauss2018,kariyappa2019}. The latter decreases the sensitivity to overfitting on patterns identified in signals. We exploit the fact that one can measure the confinement state in various different ways, reducing the dependence on specific signals, and allow for easily handling missing and/or corrupted data. Collectively, the use of different models and inputs enables for better confidence estimates by examining variability in the individual predictions~\cite{lakshminarayanan2017,rahaman2021uncertainty}.

The problem is formulated as a supervised classification task with model confidence, along with the ability to deal with missing and/or broken input signals. We incorporate neural network-based methods for exploiting sequential patterns and use decision tree-based methods for static predictions. The former further develops works on neural network-based classification for confinement states~\cite{mathews2019,zorek2022,orozco2022,gill2024,yang2024,he2024,shin2020,meakins2010,matoslhd2020,matoslhd2021}, using the Fourier Neural Operator (FNO)~\cite{li2021} with a recurrent structure~\cite{hochreiter1997}, whereas the latter is implemented with gradient-boosted decision trees (GBDT) using XGBoost~\cite{xgboost2016}. On the feature axis we define input feature sets both categorized by their approximate `domain' and combinations thereof, and use both raw diagnostic measurements and engineer physically meaningful features. The model+input combinations are fit using multiple data-splits that cover varying (mutually exclusive) groups of experimental topics to encourage model generalization. Individual configurations are empirically calibrated to ensure meaningful uncertainties, and are combined through a weighted linear combination, allowing for robust classification with uncertainty quantification.

The model is developed for the Tokamak à Configuration Variable (TCV). We aim to distinguish between the aforementioned low confinement mode (L-mode) and high confinement mode (H-mode), and an `in-between', dithering phase (D). A dataset of 302 fully labeled discharges---a confinement state label at each timestep---is used to fit and evaluate our proposed method. We publicly release the labels for this `TCV confinement state database', encouraging the community to build upon this work.

Evaluations are carried out to validate the method's prediction accuracy, the soundness of the provided confidence estimates, and the ability to deal with bad/missing data. We consider the Cohen's kappa coefficient and Expected Calibration Error as quantitative metrics covering accuracy and uncertainty. Qualitatively, we provide an extensive evaluation covering specific use-cases: ITER Baseline Scenario (IBL) plasmas~\cite{labit2024}, extrapolation to out-of-distribution regimes ($\delta_{\text{top}} > 0.3$ and $\beta_N > 1.7$), quasi-continuous exhaust (QCE) regimes~\cite{labit2019}, and unusual scenarios such as negative triangularity configurations~\cite{coda2022}. In short, our contributions can be summarized as follows:
\begin{itemize}
\item We create a dataset of confinement states for 302 TCV discharges, covering a wide variety of plasma regimes. This dataset is publicly available at~{[\textsc{released upon publication}]}.
\item We develop a method for robust confinement state classification with uncertainty quantification, using ensembles of different models and different feature sets. We use NN- and random forest-based models along with varying input sets including both raw measurements and engineered features. Through an ensembling procedure we can predict the confinement state with a meaningful prediction confidence and can deal with missing/corrupt signals.
\item We extensively evaluate the proposed method both quantitatively and qualitatively. Metrics are evaluated both for prediction accuracy and uncertainty calibration. We explore performance on a variety of plasma scenarios, consider extrapolation to out-of-distribution regimes, and extensively evaluate model robustness and the behavior of the confidence estimates.
\end{itemize}

\section{Problem Formulation}\label{sec:problem}
In a tokamak discharge, operation starts in a state which does not display significant fusion performance, which we refer to as low (L) confinement mode. Most experiments subsequently aim at transitioning to a more performing state, high (H) confinement mode. The transition between these two states has been experimentally discovered on ASDEX~\cite{hmode1982} and since then extensive studies have been conducted to explain the reasons behind this switch and the physics mechanisms involved. Nonetheless, at present, no specific set of rules exists to automatically distinguish between the two states on a large scale. The situation is rendered more difficult by the presence of intermediate states, displaying a transitional nature that makes them particularly difficult to distinguish. These states often appear around state transitions, but can also appear within L or H phases. In this context, on TCV we often observe rapid oscillations which we refer to as the Dithering (D) state~\cite{martindither2004,tcvdither2024}. 

The identification of plasma confinement states is typically carried out on a shot by shot basis by an expert. The task consists of inspecting various diagnostic signals and looking for specific signatures; for example, a trace from Edge Localized Modes (ELMs) in H$\alpha$ emissions is a potential indicator of being in H-mode. Given the difficult and time consuming nature of this process, scaling it up to large datasets becomes challenging, highlighting the utility of an automated approach.

We define the problem setting of automated detection of the confinement state as learning a function $f$ that maps measured and computed plasma quantities to a prediction of the confinement state. This is described as follows:
\begin{align}
f: \mathbf{x}^{\mathbf{u},\mathbf{t}^{\textit{in}}} \in \mathbb{R}^{\textit{U} \times \textit{T}^{\textit{in}}} \rightarrow \mathbf{y}^{\mathbf{s},\mathbf{t}^{\textit{out}}} \in \mathbb{R}^{3 \times \textit{T}^{\textit{out}}}, 
\label{eq:formulation}
\end{align}
where $\mathbf{x}^{\mathbf{u},\mathbf{t}^{\textit{in}}}$ represents the input matrix with $\textit{U}$ input signals and $\textit{T}^{\textit{in}}$ time samples, and $\mathbf{y}^{\mathbf{s},\mathbf{t}^{\textit{out}}}$ represents the output prediction matrix of states $\mathbf{s} = \{L, D, H\}$ for $\textit{T}^{\textit{out}}$ time samples. Note that $\textit{T}^{\textit{in}}$ and $\textit{T}^{\textit{out}}$ need not be the same; in practice we do not predict at each measurement for efficiency reasons, i.e. $\textit{T}^{\textit{out}} < \textit{T}^{\textit{in}}$.

For each timestep, the output predictions sum to one and are nonnegative, i.e.
\begin{align}
    \forall_{t \in \mathbf{t}^{\textit{out}}}\textstyle\sum_{s\in \mathbf{s}}y^{s,t} = 1, \hspace{.5cm} \forall_{t \in \mathbf{t}^{\textit{out}},s \in \mathbf{s}} \ y^{s,t} > 0,
\end{align}
Consequently, we can interpret our function as the conditional distribution between the input measurements and the confinement state, $p(y|x)$. We then define our prediction confidence (i.e. certainty) as the probability of the maximum class:
\begin{align}
    \textit{confidence} = \max_{s \in \mathbf{s}}p(y = s | x),
    \label{eq:confidence}
\end{align}
This quantity should be calibrated, e.g., if it is 0.8, we expect---on a sufficiently large sample---a prediction accuracy of 80\%.

Additionally, we assume that not all input signals in $\mathbf{x}$ are always present and perfectly acquired. Over time there are periods where a diagnostic is not available, or instances where it failed to acquire correctly. It is crucial that the method is robust to these failure modes to ensure the applicability to as many experiments as possible.  %

\section{Dataset}\label{sec:data}

We present a database of 302 TCV discharges labeled with confinement states `Low', `Dithering', or `High' over their entire duration. Discharges were selected to cover a large variety of confinement behaviors: H-modes with steady edge-localized modes (ELMs), ELM-free regimes, dithering phases near the transition threshold, H-L back transitions, among others. The dataset includes experiments from various missions, such as studies on the ITER baseline scenario, disruption avoidance, L-H transitions, density limits, and the effects of heating methods and plasma shaping. Labeling and data extraction was handled using the DEFUSE framework~\cite{defuseiaea}.

\subsection{Confinement State Labeling}

\begin{figure}[t]
\begin{center}\includegraphics[width=.75\linewidth]{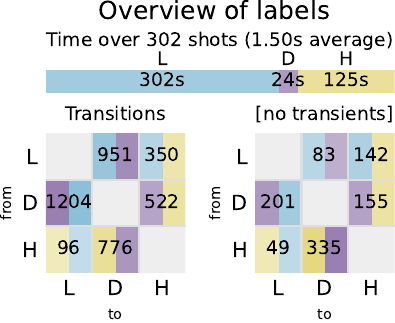}\end{center}
    \caption{Overview of the total time and transitions present in the dataset. The top shows the cumulative time spent in the different states, whereas the bottom depicts the total number of state transitions. The left depicts the total number of transitions, whereas the right excludes unstable or transient transitions, which we define as those where the plasma changes state within \SI{10}{\milli\second} before or after a transition.}
    \label{fig:dataset_labels}%
\end{figure}

\begin{figure}[t]
\begin{center}\includegraphics[width=1\linewidth]{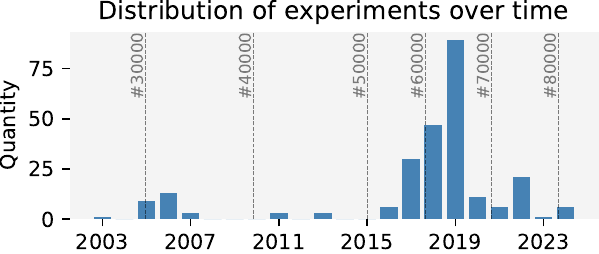}\end{center}
    \caption{The distribution of the dates of the discharges present in the dataset. It includes TCV plasmas from several decades, with the  majority from the last 10 years.}
    \label{fig:dataset_when}%
\end{figure}

\begin{figure}[t]
\begin{center}\includegraphics[width=1\linewidth]{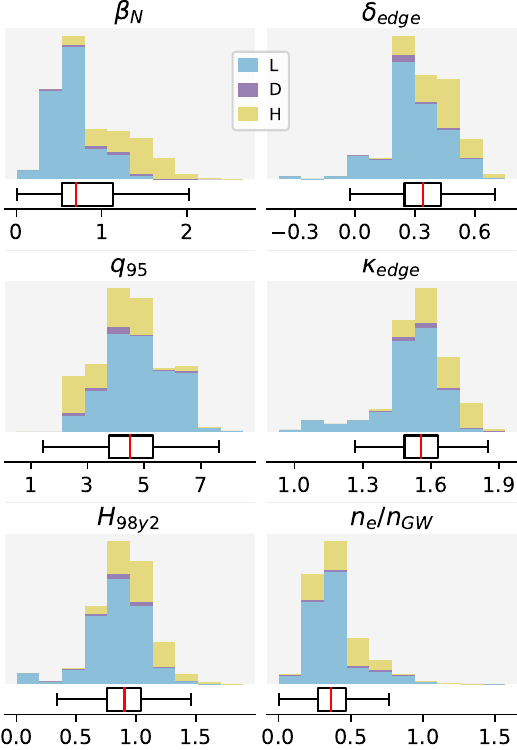}\end{center}
    \caption{Distributions of key plasma parameters in the dataset, stacked for L, D and H-mode plasmas. We plot average values for phases of \SI{20}{\milli\second}--around the TCV energy confinement time--to exclude fast transient measurements.}
    \label{fig:dataset_eda}%
\end{figure}

In total, the dataset covers \SI{451.67}{\second} of plasma dynamics, an average time of \SI{1.50}{\second} per discharge. Of this time, \SI{302.20}{\second} is spent in L-mode, \SI{24.08}{\second} in dithering and \SI{125.38}{\second} in H-mode, see Figure~\ref{fig:dataset_labels} for a depiction of the time distribution and the transitions between the different states. States are labeled at a precision of \SI{10}{\kilo\hertz}, giving a total of approximately 4.5 million timeslices. The dataset covers experiments between 2003 and 2024, see Figure~\ref{fig:dataset_when} for the distribution of shots over time. To illustrate the types of plasma scenarios we plot the distribution of key parameters in Figure~\ref{fig:dataset_eda}.

The labeling was done by a single expert to ensure consistency throughout the whole dataset; we found that inconsistent labeling is one of the biggest detractors to model performance~\cite{matoslhd2021}. 
For this annotation, one of the main identifiers of a change in confinement is the plasma emission, which is generally visible in the photodiode signal. Kinetic profiles are also a key indicator, albeit generally less available; contextual quantities such as the plasma stored energy are also used. In certain marginal scenarios however, it can be difficult to label the state with high certainty. We aim for consistency in these scenarios to maximize model performance, however potential biases of the human expert cannot be avoided in the supervised learning setting.

\subsection{Signals}
\setlength\tabcolsep{1pt}
\begin{table*}[h]
\footnotesize
\begin{tabular}{p{0.4999999\textwidth}p{0.4999999\textwidth}}
\vtop{\begin{minipage}{0.4999999\textwidth}
\vspace{0.5em}\begin{center}\textbf{Shaping}\end{center}\vspace{-1.2em}
\flushleft
\begin{tabular}{p{0.22\textwidth}p{0.12\textwidth}p{0.6\textwidth}}
\toprule
Variable & Unit & Description \\
\midrule
$A_p$ & \SI{}{\meter\squared} & Plasma cross-sectional area \\
$\delta_{\text{bottom}}$ & \hphantom{.} & Lower (bottom) plasma triangularity \\
$\delta_{\text{top}}$ & \hphantom{.} & Upper (top) plasma triangularity \\
$\Delta_{\text{in}}$ & \SI{}{\meter} & Radial gap between plasma edge and inner wall \\
$\Delta_{\text{out}}$ & \SI{}{\meter} & Radial gap between plasma edge and outer wall \\
$\kappa$ & \hphantom{.} & Plasma elongation \\
$R_0$ & \SI{}{\meter} & Major radius of the plasma \\
$a$ & \SI{}{\meter} & Minor radius of the plasma \\
$R_\text{axis}$ & \SI{}{\meter} & Radial coordinate of the magnetic axis \\
$Z_\text{axis}$ & \SI{}{\meter} & Vertical coordinate of the magnetic axis \\
$V_p$ & \SI{}{\meter\cubed} & Plasma volume \\
\bottomrule
\end{tabular}
\end{minipage}} & \vtop{\begin{minipage}{0.4999999\textwidth}
\vspace{0.52em}\begin{center}\textbf{Emissions}\end{center}\vspace{-1.05em}
\flushright
\begin{tabular}{p{0.14\textwidth}p{0.1\textwidth}p{0.7\textwidth}}
\toprule
Variable & Unit & Description \\
\midrule
$\text{PD}_{\textit{H}\alpha}^{}$ & \SI{}{\volt} & Photodiode (PD) signal for \textit{H}$\alpha$/\textit{D}$\alpha$ line emission ($\lambda$=656.3 nm) \\
$\text{PD}_{\textit{CIII}}^{}$ & \SI{}{\volt} & Photodiode signal for CIII line emission ($\lambda$=465.1 nm) \\
$\text{PD}_{\text{FFT}}^{}$ & $\text{a.u.}$ & Spectral features from PD signals, computed as the variance of frequency spectra over sliding windows \\
\bottomrule
\end{tabular}

\centering
\vspace{0.3em}\begin{center}\textbf{Magnetics}\end{center}\vspace{-1.2em}
\flushright
\begin{tabular}{p{0.14\textwidth}p{0.1\textwidth}p{0.7\textwidth}}
\toprule
Variable & Unit & Description \\
\midrule
$B_0$ & $\SI{}{\tesla}$ & Vacuum toroidal magnetic field at $R=\SI{0.88}{\meter}$ \\
$I_{p}$ & $\SI{}{\ampere}$ & Plasma current \\
$I_{p,\textit{ref}}$ & $\SI{}{\ampere}$ & Prescribed plasma current \\
$q_{95}$ & \hphantom{.} & Safety factor at 95\% of enclosed magnetic flux \\
\bottomrule
\end{tabular}

\end{minipage}} \\
\multicolumn{2}{c}{
\begin{minipage}{1\textwidth}
\vspace{0.2em}\begin{center}\textbf{Density}\end{center}\vspace{-2.2em}
\flushleft
\begin{tabular}{p{0.18\textwidth}p{0.18\textwidth}p{0.63\textwidth}}
\toprule
Variable & Unit & Description \\
\midrule
$n_{e,\text{core}}$ & \SI{}{\per\meter\squared} & Vertical interferometer line-integrated electron density from 0.87 m < ch < 0.91 m \\
$n_{e,\text{LFS}}$ & \SI{}{\per\meter\squared} & Vertical interferometer line-integrated electron density from ch > 1.03 m \\
$n_e/n_{\textit{GW}}$ & \hphantom{.} & Greenwald fraction~\cite{Greenwald2002} using electron density measurements from interferometry \\
$\text{max}(n'_{e,\text{edge}})$ & \SI{}{\per\meter\cubed\per\rho} & Maximum first derivative of $n_{e,\rho}$ in edge region ($0.85 < \rho < 0.95$) from Thomson Scattering \\
$\text{max}(n''_{e,\text{edge}})$ & \SI{}{\per\meter\cubed\per\rho\squared} & Maximum second derivative of $n_{e,\rho}$ in edge region ($0.85 < \rho < 0.95$) from Thomson Scattering \\
$n_{e,0}$ & \SI{}{\per\meter\cubed} & On-axis ($\rho=0$) electron density from Thomson Scattering \\
\bottomrule
\end{tabular}
\end{minipage}}  \\ 
\multicolumn{2}{c}{
\begin{minipage}{1\textwidth}
\vspace{1.4em}\begin{center}\textbf{Temperature}\end{center}\vspace{-2.2em}
\flushleft
\begin{tabular}{p{0.18\textwidth}p{0.18\textwidth}p{0.63\textwidth}}
\toprule
Variable & Unit & Description \\
\midrule
$\textit{SXR}_{\text{core}}$ & \SI{}{\watt\per\meter} & Soft X-Ray core ($\rho < 0.15$) emission \\
$\text{max}(T'_{e,\text{edge}})$ & \SI{}{\electronvolt\per\rho} & Maximum first derivative of $T_{e,\rho}$ in edge region ($0.85 < \rho < 0.95$) from Thomson Scattering \\
$\text{max}(T''_{e,\text{edge}})$ & \SI{}{\electronvolt\per\rho\squared} & Maximum second derivative of $T_{e,\rho}$ in edge region ($0.85 < \rho < 0.95$) from Thomson Scattering \\
$T_{e,0}$ & \SI{}{\electronvolt} & On-axis ($\rho=0$) electron temperature from Thomson Scattering \\
\bottomrule
\end{tabular}
\end{minipage}} \\
\begin{minipage}{0.4999999\textwidth}
\vspace{1.55em}\begin{center}\textbf{Power}\end{center}\vspace{-1.2em}
\flushleft
\begin{tabular}{p{0.24\textwidth}p{0.12\textwidth}p{0.6\textwidth}}
\toprule
Variable & Unit & Description \\
\midrule
$P_{\textit{in}}$ & $\SI{}{\watt}$ & Total input power \\
$P_{\textit{OHM}}$ & $\SI{}{\mega\watt}$ & Ohmic heating power \\
$P_{\textit{NBI}}$ & $\SI{}{\mega\watt}$ & Delivered NBI power \\
$P_{\textit{NBI2}}$ & $\SI{}{\mega\watt}$ & Delivered NBI2 power \\
$P_{\textit{ECRH}}$ & $\SI{}{\mega\watt}$ & ECRH power\\
$P_{\textit{LH}}$ & \hphantom{.} & LH power threshold scaling from~\cite{hmodeiter2008}\vspace{0.3em} \\
\bottomrule
\end{tabular}
\end{minipage}  & 
\begin{minipage}{0.4999999\textwidth}
\vspace{1.35em}\begin{center}\textbf{Energy Content}\end{center}\vspace{-1.2em}
\flushright
\begin{tabular}{p{0.14\textwidth}p{0.09\textwidth}p{0.71\textwidth}}
\toprule
Variable & Unit & Description \\
\midrule
$\beta_{N}$ & \hphantom{.} & Normalized toroidal beta ($\beta_{t}\frac{aB_0}{I_p}$)\\
$\beta_{p}$ & \hphantom{.} & Poloidal beta \\
$\beta_{t}$ & \hphantom{.} & Toroidal beta\\
${W}_{\textit{tot}}$ & $\SI{}{\joule}$ & Total plasma stored energy \\
$\text{DML}$ & \SI{}{\weber} & Plasma toroidal flux from the diamagnetic loop \vspace{0.01cm}\\
$H_{\textit{98y2}}$ & \hphantom{.} & Energy confinement time normalized to $\tau_{E}^{\text{IPB98(y, 2)}}$~\cite{Transport1999} \\
\bottomrule
\end{tabular}
\end{minipage}
\\
\begin{minipage}{0.4999999\textwidth}
\vspace{1.45em}\begin{center}\textbf{Radiation}\end{center}\vspace{-1.2em}
\flushleft
\begin{tabular}{p{0.24\textwidth}p{0.12\textwidth}p{0.6\textwidth}}
\toprule
Variable & Unit & Description \\
\midrule
$P_{\textit{rad}}$ & $\SI{}{\kilo\watt}$ & Total radiated power from bolometers \\
$P_{\textit{rad},\text{bulk}}$ & $\SI{}{\kilo\watt}$ & Bulk ($\rho < 1$) radiated power from bolometers \\
$P_{\textit{rad},\text{SOL}}$ & $\SI{}{\kilo\watt}$ & Scrape-Off Layer radiated power from bolometers \vspace{0.115cm}\\
\bottomrule
\end{tabular}
\end{minipage}
 &
\begin{minipage}{0.4999999\textwidth}
\centering
\vspace{1.5em}\begin{center}\textbf{Other}\end{center}\vspace{-1.2em}
\flushright
\begin{tabular}{p{0.14\textwidth}p{0.09\textwidth}p{0.71\textwidth}}
\toprule
Variable & Unit & Description \\
\midrule
$l_i$ & \hphantom{.} & Internal inductance of the plasma current \\
$Z_{\textit{eff}}$ & \hphantom{.} & Effective ion charge \\
$\nu_{e,\text{ped}}^{*}$ & \hphantom{.} & Normalized edge electron collisionality~\cite{labit2021iaea} \\
$V_{\textit{loop}}$ & $V$ & Loop voltage\\
\bottomrule
\end{tabular}
\end{minipage}
\end{tabular}
\caption{The list of input signals and constructed features used in the proposed method to classify the confinement state of TCV. Signals and features are grouped by diagnostic systems or physical quantities. Shaping and energy content-related features originate from LIUQE~\cite{moret2015liuqe}.}\label{tab:signals}
\end{table*}
\setlength\tabcolsep{6pt}

In this work, we utilize a broad set of signals to automatically identify the confinement state. These signals are selected to measure plasma quantities in a variety of ways, adding redundancy to increase robustness and reliability. We split them into a set of categories that group them by diagnostic systems or physical quantities. The categorization we consider consists of \textit{shaping, emissions, magnetics, density, temperature, power, energy content, radiation}, and a miscellaneous \textit{other} category. An overview of all signals and the categorization is provided in Table~\ref{tab:signals}.

All signals are interpolated to a common timebase of \SI{10}{\kilo\hertz} using linear interpolation. For real-time applications one must use causal interpolation, but given that the scope of this work is offline analysis, we use linear interpolation to maximize information at each timestep. For more details on how these signals are used as inputs for the different models, we refer to Section~\ref{ss:components}.

\subsection{Dataset availability}
 The dataset is publicly available at~{[\textsc{released upon publication}]}

\section{Method}\label{sec:method}
\begin{figure*}[t]
\begin{center}\includegraphics[width=.94\linewidth]{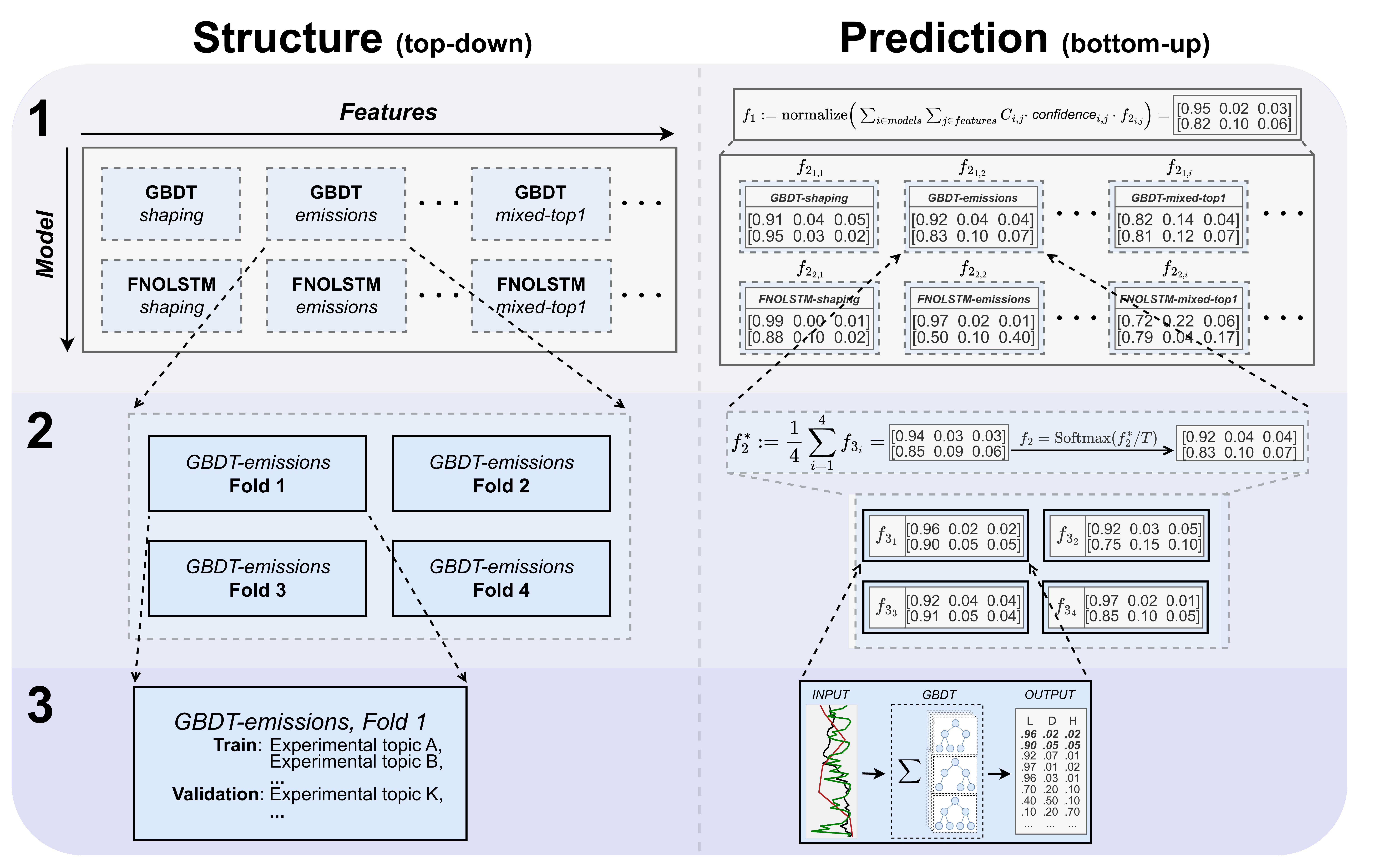}\end{center}
    \caption{An overview of the ensemble structure and prediction procedure. The structure is defined in a top-down fashion. First, we ensemble over different models and feature sets (level 1). Each \textit{(model + feature set)} combination consists of a small ensemble of models fit on different folds of the data (level 2). At the bottom, we consider individual models (level 3). Prediction is done in a bottom-up fashion. Individual models map the input signals to a prediction of being in L, D or H (level 3). These predictions are combined and re-scaled on the level of a \textit{(model + feature set)} (level 2). The resulting predictions are combined weighted by their confidences and a constant that is computed using a \textit{(model + feature set)}'s average classification performance, giving the full ensemble prediction (level 1).
    }\label{fig:ensemble}%
\end{figure*}

The overarching goal of our approach is to maximize classification performance under two constraints: (1) providing meaningful, ideally calibrated, uncertainties, and (2) being robust to missing or corrupted input signals. To do so, we propose ensembling a set of models on two axes: different model formulations and different feature sets. In this section we discuss this setup and the components in detail; an overview of the method is provided in Figure~\ref{fig:ensemble}.

\subsection{Structure}\label{ss:components}
We describe our ensembling procedure in a top down manner using three `levels' in a hierarchy. This structure is depicted in Figure~\ref{fig:ensemble} on the left. The three levels are as follows:
\begin{enumerate}[label=(\arabic*)]
    \item Ensembling over two axes: model formulations and feature sets. A single element of the ensemble is defined as a \textit{(model + feature set)}.
    \item For each \textit{(model + feature set)}, we ensemble over different train and validation splits (folds). The splits are chosen to have no overlap in the validation sets w.r.t. experimental topics to encourage variety and generalization. 
    \item The lowest level describes a single model, with a single feature set, fit on a single fold of the dataset.
\end{enumerate}
The ensembling steps on levels 1 and 2 serve the function of providing variety in the model behavior, by altering the formulations, feature sets and data splits. Generally, such an approach contributes to better uncertainty estimates. Intuitively, if diverse models agree on a label, the prediction is unlikely to be an artifact of an individual model, allowing us to assign higher confidence, see e.g.~\cite{lakshminarayanan2017,rahaman2021uncertainty} for more details.

Additionally, the use of different feature sets addresses the point of robustness: by having models trained on many different subgroups of features, the method naturally becomes resilient to corrupted signals since they are only present in a subset of the predictors, and any given feature never appears in all feature sets. This advantage extends to dealing with missing signals for models that require all inputs to be present: since not all models utilize all signals, we can still use a subset of models when some signals are missing. Effectively, we can utilize many signals if they are available, but we do not require them.

\subsection{Components}\label{sec:components}
The axes of variation in our approach are the feature sets and the model inductive biases. In this section, we discuss the construction of these feature sets, and describe the models in detail. Specifically, for the latter, we consider a dynamic and static formulation, based on neural networks (FNOLSTM), and tree ensembles (GBDT), respectively.

\textbf{Feature sets.}
We employ the features introduced before in Table~\ref{tab:signals}. The individual feature sets are constructed either by taking features only from one category, or by mixing features from all categories. When taking a subset of a category, we first order them by their individual discriminative power, which is computed by fitting simple models on single features, see~\ref{ap:feature_sets} for more details. For example, if we only take two features from category `Shaping', we take the two most informative according to a precomputed metric, rather than picking them arbitrarily. 

Given this scheme, we construct two groups for each category: one with the top-$k$ features, and one with all features. Additionally, we construct various groups mixing all categories, taking the top-1, top-2, etc.; see~\ref{ap:feature_sets} for all \textit{(model + feature set)} combinations. 

The main motivation for this approach is twofold. For one, by constructing sets in an informed manner, we can fit multiple models covering the main aspects of the plasma whilst having little overlap in their inputs. Ideally, this strategy gives us distinct models with good performance,  reducing the sensitivity to single features. Additionally, having models cover a single category provides a degree of interpretability, given that we can inspect individual model predictions.

\textbf{Dynamic model formulation: FNOLSTM.}
First, we consider the problem in a dynamic formulation. The model maps an input sequence of signal data, up to a given timestep $t_m$, to the label at time $t_{m-k}$, i.e., a small offset $k$ before the last input. This formulation can be expressed as follows:
\begin{align}
f_{\text{dynamic}}: \mathbf{x}^{\mathbf{u},\mathbf{t}_{\leq m}} \in \mathbb{R}^{\textit{U} \times m}\rightarrow \mathbf{y}^{\mathbf{s},t_{m-k}}\in \mathbb{R}^{3},
\label{eq:fdynamic}
\end{align}
with the same notation as Equation~\ref{eq:formulation}, and $f_{\text{dynamic}}$ denotes the input-output map we learn. Intuitively, we assume knowledge of all signal data up to time $t_m$, and provide predictions with a lag of $k$ timesteps.

To implement $f_{\text{dynamic}}$ we use artificial neural networks (NNs) to best exploit the potentially subtle information carried by the input dynamics. Previous works have shown success on confinement state classification with NNs~\cite{mathews2019,zorek2022,orozco2022,gill2024,yang2024,he2024,shin2020,meakins2010,matoslhd2020,matoslhd2021}. We build upon the general principle of feature extraction on small timescales using convolution-like methods and on large timescales using recurrent methods.

We use the Fourier Neural Operator (FNO)~\cite{li2021} on small input windows of signals. The FNO combines a linear, global integral operator with non-linear, local activation functions, which has proven highly successful on modeling the dynamics of various physical processes~\cite{wen2022,thorsten2023,Poels2023neuralPDE,Gopakumar2024PlasmaSurrogate}. Specifically, the FNO performs a Fast Fourier Transform (FFT)~\cite{cooley1965} of the input signals, in our case along the time axis, after which we perform a matrix multiplication on the spectral coefficients. The result is transformed back and summed to a point-wise transformation of the grid. This procedure can be expressed as follows:
\begin{align}
    \textit{FNO}^l: \mathbf{z}^l = \sigma\big(\text{FFT}^{-1}(\mathbf{R}^l\text{FFT}(\mathbf{z}^{l-1})) + \mathbf{W}^l \mathbf{z}^{l-1}\big),
    \label{eq:fno}
\end{align}
mapping an input (multidimensional) signal at layer $l-1$ ($\mathbf{z}^{l-1}$) to the output at layer $l$ ($\mathbf{z}^{l}$). The learned weight matrices are denoted as $\mathbf{R}^l \in \mathbb{R}^{D \times D \times M}$ and $\mathbf{W}^l \in \mathbb{R}^{D \times D}$, for $D$ hidden dimensions and $M$ fourier modes, with non-linear activation function $\sigma$.

\begin{figure*}[t]
\begin{center}\includegraphics[width=.7\linewidth]{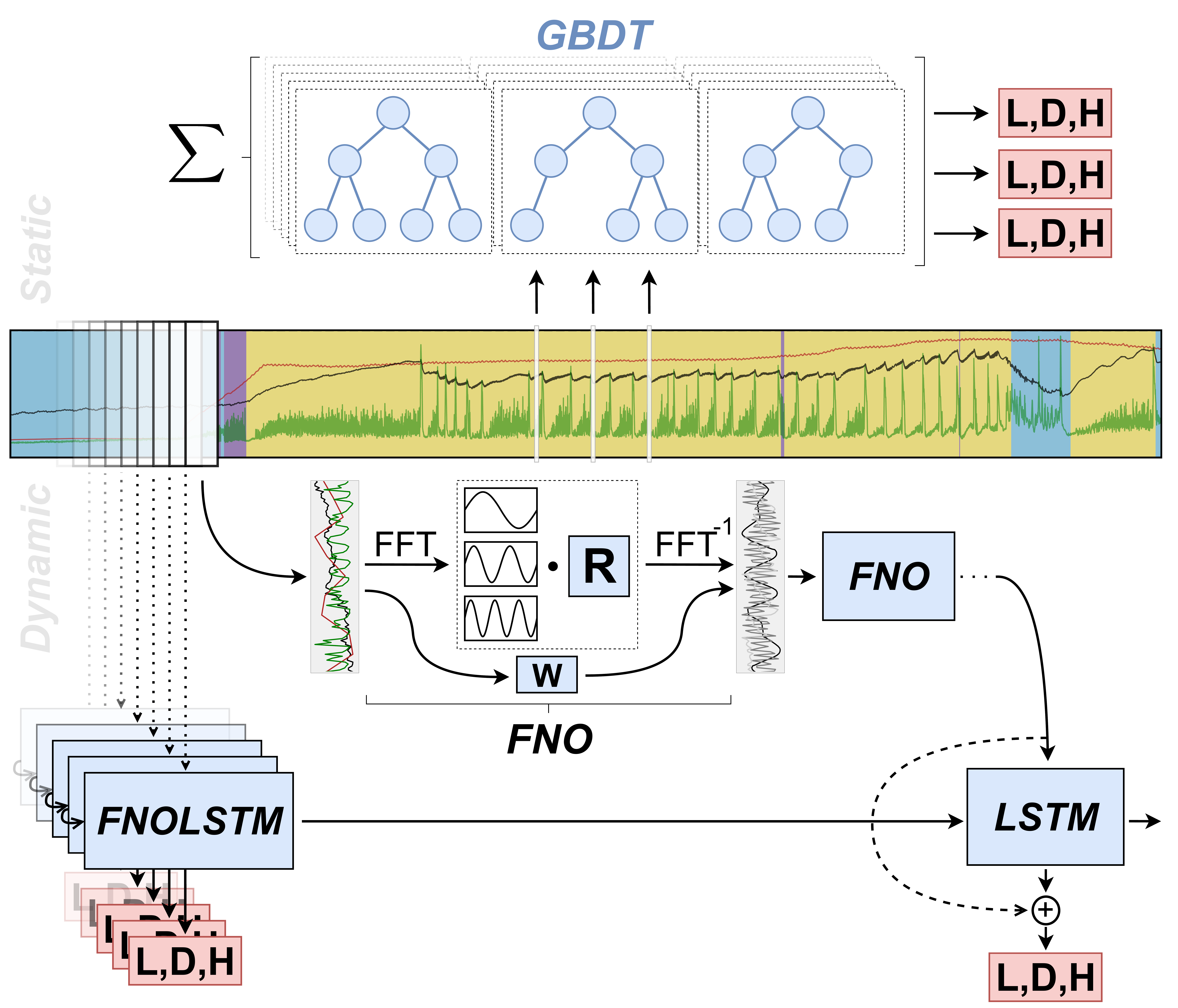}\end{center}
    \caption{A simplified illustration of the two modeling approaches, a static formulation operating on timeslices of signals (top) and a dynamic formulation operating on sequences of signals (bottom). \textbf{Static (GBDT):} The static formulation is implemented with gradient boosted decision trees. A collection of small decision trees is fit in a sequential manner: each tree is fit using the residual errors of the previously fit trees. These trees operate on a vector of signal data corresponding to a single timeslice of the discharge. The final prediction is computed using a weighted combination of all individual trees. \textbf{Dynamic (FNOLSTM):} The dynamic formulation is implemented through a neural network extracting features on both small and large timescales. The input consists of a matrix corresponding to a small time window of signal data. This input is transformed with the FNO, a convolution-like operator acting primarily in the frequency domain. This computed abstract representation is processed in a recurrent manner over the entire duration of the shot with an LSTM. The final prediction is then computed using this local and global representation of the input signals.}
    \label{fig:fnolstm}%
\end{figure*}

To capture dynamics on longer time scales, we combine the local feature extractor with a recurrent architecture that operates on sequences of arbitrary length. We utilize the Long Short Term Memory (LSTM)~\cite{hochreiter1997} architecture, which can be summarized as follows:
\begin{align}
    \textit{LSTM}: \mathbf{h}_{t_m}, \mathbf{c}_{t_m} = f_{\textit{LSTM}}(\mathbf{z}_{t_m},\mathbf{h}_{t_{m-1}},\mathbf{c}_{t_{m-1}}),
    \label{eq:lstm}
\end{align}
where $f_{\textit{LSTM}}$ denotes the learned recurrent layer, $\mathbf{h}_{t_m}$ and $\mathbf{c}_{t_m}$ the hidden state components of the LSTM unit, and $\mathbf{z}_{t_m}$ the input of the recurrent layer, that is, the output of Equation~\ref{eq:fno}. Note the temporal connection of the LSTM: it takes as input the previous hidden state ($\mathbf{h}_{t_{m-1}}, \mathbf{c}_{t_{m-1}}$) along with the current signal information $\mathbf{z}_{t_m}$ to compute the outputs at time $t_m$.

The output of the LSTM is passed through a small set of fully connected neural network layers, i.e. a small Multi-Layer Perceptron (MLP)~\cite{ivakhnenko1971}. The resulting value is subsequently mapped to probabilities using the Softmax function, giving us final prediction $\mathbf{y}$. Additionally, we add a skip connection~\cite{he2016} between the local feature extractor and the input of the MLP. In certain instances the input time window in isolation already sufficiently describes the plasma state: the skip connection removes the potential bottleneck of the hidden state.

Lastly, we discuss the specifics of the inputs and outputs. The input at time $t_m$ is a time window ending at time $t_m$ of size $w$ timesteps, that is, covering the interval $(t_{m-w}, t_m]$, or denoted as input matrix $\mathbf{x}_{t_{m-w}:t_m} \in \mathbb{R}^{U \times w}$. The model operates on a stride of $p$: we predict every $p$ timesteps. Finally, the prediction is given with a lag of $k$ timesteps in order to maximize accuracy---in an online setting, one could imagine minimizing this delay $k$, or setting it to $k=0$ for no latency. Altogether, this process is described as follows:
\begin{align}
\mathbf{z}_{t_m} &= \textit{FNO}(\mathbf{x}_{t_{m-w}:t_m}), \\
\mathbf{h}_{t_m}, \mathbf{c}_{t_m} &= \textit{LSTM}(\mathbf{z}_{t_m}, \mathbf{h}_{t_{m-p}}, \mathbf{c}_{t_{m-p}}),\\
f_{\text{dynamic}} : \mathbf{y}^{\mathbf{s},t_{m-k}} &= \text{Softmax}\big(\textit{MLP} \big(\mathbf{z}_{t_m} + \mathbf{h}_{t_m}\big)\big),
\label{eq:fdynamic_fnolstm}
\end{align}
for the dynamic model $f_{\text{dynamic}}$, and $\mathbf{x}$ and $\mathbf{y}$ as in Equation~\ref{eq:fdynamic}. A simplified illustration is provided in Figure~\ref{fig:fnolstm} (bottom).

\textbf{Static model formulation: GBDT.}
Secondly, we consider the problem in a static formulation. Here, the model maps inputs at a given timestep $t_m$ to the corresponding label at this time, i.e.,
\begin{align}
f_{\text{static}}: \mathbf{x}^{\mathbf{u},t_m} \in \mathbb{R}^{\textit{U}}\rightarrow \mathbf{y}^{\mathbf{s},t_m}\in \mathbb{R}^{3},
\label{eq:fstatic}
\end{align}
with the same notation as Equation~\ref{eq:formulation}, and $f_{\text{static}}$ denoting the input-output map we learn.

We implement $f_{\text{static}}$ using gradient boosted decision trees (GBDT)~\cite{friedman2001greedy}, a machine learning method that builds an ensemble of small decision trees in a sequential manner. GBDTs have shown strong performance on tabular data~\cite{vadim2024,kit2023}, making them well suited for the static formulation.

Each individual tree splits an input sample iteratively based on feature thresholds up to a leaf node holding the prediction value. A benefit of this formulation is the ability to deal with missing signals by simply choosing a default direction on a split node. GBDTs build an ensemble of decision trees in a sequential manner, by fitting individual trees using residual errors of previous trees w.r.t. the train dataset. This iterative process is carried out through gradient boosting with XGBoost~\cite{xgboost2016}, i.e., the residuals are computed as the gradient of a specified loss function.

In practice, we fit an ensemble of trees for each {$\text{class} \in \{L, D, H\}$}. The final prediction for the GBDT model is obtained by normalizing over the three ensemble predictions using Softmax. This entire process is described as follows:
\begin{align}
\begin{split}
    f_{\text{static}}: \mathbf{y}^{\mathbf{s},t_m} = \text{Softmax}\big(&\sum_{k=1}^K f_{c,k} (\mathbf{x}^{\mathbf{u},t_m}) \\
    &{\text{ for } c \in \{L, D, H\}}\big),
\end{split}
\label{eq:fstatic_gbdt}
\end{align}
where $K$ denotes the number of trees in an individual class' ensemble, $f_{c,k}$ an individual decision tree; $\mathbf{x}$ and $\mathbf{y}$ as in Equation~\ref{eq:formulation}. A simplified illustration of this process is given in Figure~\ref{fig:fnolstm} (top).

\subsection{Fitting Procedure}
We describe the fitting procedure in a bottom-up manner, starting from the individual models up to the full ensemble (i.e. from level 3 to 1). An overview of this process is depicted in Figure~\ref{fig:ensemble} (right). The dataset is split into a subset for fitting the individual models (`train-validation'), a subset for fitting ensembling parameters (`ensemble-holdout'), and the test set used in Section~\ref{sec:experiments}. Since the method should be applicable to novel scenarios, we choose the splits to minimize overlap in the contained missions. This approach ensures we evaluate on sufficiently different scenarios, avoiding information leakage.

\textbf{Individual model training.}
Each individual \textit{(model + feature set)} consists of a small ensemble that shares the same model hyperparameters (level 2). Each element of this ensemble, an individual model (level 3), is fit using a different split of the `train-validation' set, which is referred to as a fold. See Figure~\ref{fig:ensemble} for a visualization.

For each fold, we optimize a separate set of model parameters using the confinement state labels. For both the FNOLSTM and the GBDT-based models, the loss is categorical cross-entropy between the model outputs and the ground-truth labels. For the FNOLSTM, given its sequential nature, we optimize using subsequences of discharges, which are sampled to ensure a balance in the output labels, see~\ref{ap:model_training} for more details. For the GBDT, given its static nature, we sample individual timeslices. Lower-frequency states are oversampled for better balance, and periods around state transitions are oversampled to ensure we capture their dynamics, see~\ref{ap:model_training} for details.

\textbf{Ensembling procedure.}
The prediction procedure can be split into three steps. First, we predict individually with each model (level 3), i.e. Equation~\ref{eq:formulation}. These predictions are consequently averaged at the level of the \textit{(model + feature set)}, level 2. This mini-ensemble is calibrated with the `ensemble-holdout' set using temperature scaling~\cite{guo2017}, which is defined as follows:
\begin{align}
f_{2}^* &= \frac{1}{N} \sum_{i=1}^{N} f_{3_i}, \\
f_2 &= \text{Softmax}(f^{*}_{2}/T),
\end{align}
for level-2 ensemble $f_2$, taking as input $N$ individual model outputs $f_{3_i}$, and $T$ denoting the fitted temperature parameter. Prior work shows that this pool-then-calibrate approach generally results in the lowest calibration error~\cite{rahaman2021uncertainty}. 

The level 1 prediction, the final output, consists of a linear combination of all individually calibrated level 2 outputs. Each output is weighted by a constant weight, denoted as $C_{i,j}$, for $i \in models$ and $j \in features$. These constants are determined using the classification performance. Specifically, we compute the Cohen's kappa coefficient~\cite{cohen1960} (see Section~\ref{sec:metrics}) for more information), which we normalize over all the different models, and then square the results. This rescaling puts the constants $C_{i,j} \in [0, 1]$, while placing more emphasis on the best performing models. The final prediction then sums over all level 2 predictions, scaled by these constants and the prediction confidences (Equation~\ref{eq:confidence}), and is subsequently normalized such that the total sums to 1:

\begin{align}
\begin{split}
f&_{1} = \\
&\text{normalize}\Big( \sum_{\substack{i \in models \\ j \in features}} C_{i,j} \cdot \textit{confidence}_{i,j} \cdot f_{2_{i,j}}\Big),
\end{split}
\end{align}
for full ensemble $f_1$. An overview of the full prediction procedure is provided in Figure~\ref{fig:ensemble} (right).

\section{Experiments and Results}\label{sec:experiments}

In this section we evaluate the proposed method both with regards to its accuracy for labeling and the soundness of the confidence estimates. We consider both aggregate statistics and zoom in on specific scenarios. Challenging scenarios are included to further assess the method. Additionally, we evaluate the method with conditionally averaged behavior around specific types of transitions.

A short summary of the training specifics, including hyperparameters and the dataset splits, is provided in Section~\ref{sec:hyperparam}. We follow with quantative and qualitative evaluations in Sections~\ref{sec:metrics} and~\ref{sec:qualitative} respectively. Next, we consider extrapolation and robustness in Section~\ref{sec:extrapolation_robustness}, and conclude with conditionally averaged behavior of the confidence estimates in Section~\ref{sec:conditional_average}.

\subsection{Dataset split and hyperparameters}\label{sec:hyperparam}
\textbf{Dataset split.} To recap, the dataset (302 shots) is split into the `train-validation' set (258 shots), the `ensemble-holdout' (10 shots) and the test set (34 shots). In this section, unless mentioned otherwise, we only show results from the test set. We carefully construct the `ensemble-holdout' and test set such that they are a representative sample of the operational space of TCV. To do so, we sample these shots to approximate the distribution of the campaigns present in the full dataset. We highlight shots from plasmas presented in~\cite{labit2024} to cover the ITER Baseline (IBL). Furthermore, we select sets of shots from underrepresented scenarios to evaluate the method under more challenging circumstances. Specifically, we include shots from quasi-continuous exhaust (QCE) regimes~\cite{labit2019}, negative triangularity (NT) configurations~\cite{coda2022}, and filter specifically on $\delta_{\text{top}} > 0.3$ and $\beta_N > 1.7$ to evaluate out-of-distribution regimes.

\textbf{Hyperparameters.} In total, we utilize an ensemble of 52 models (level 1), 26 based on the FNOLSTM and 26 on GBDT. For all \textit{(model + feature set)} combinations, we refer to~\ref{ap:feature_sets}. For each \textit{(model + feature set)} configuration (level 2), we use 4 different folds. Hyperparameters are shared on this level, whereas each model is naturally defined by its own parameters. All models are optimized using the `train-validation' set. Individual model parameters are fit on the train set, whereas hyperparameters are optimized using Bayesian optimization on the validation set. The optimization target is the Cohen's kappa coefficient (see Section~\ref{sec:metrics}) averaged over the different folds. For details on the hyperparameter ranges we refer to~\ref{ap:parameters}.

All neural network models are implemented using PyTorch~\cite{paszke2019}. The gradient-boosted decision trees are implemented using XGBoost~\cite{xgboost2016}. We utilize Optuna~\cite{optuna2019} for the hyperparameter optimization and net:cal~\cite{kuppers2020} for calibrating level 2 ensembles through temperature scaling.

\subsection{Quantitative results for accuracy and calibration}\label{sec:metrics}
\textbf{Metrics.} For quantitative evaluation, we consider Cohen's kappa coefficient~\cite{cohen1960} for the classification performance, and the Expected Calibration Error (ECE)~\cite{degroot1983,naeini2015} for the calibration of the model confidence outputs. 

\textit{Cohen's kappa coefficient}~\cite{cohen1960} measures the agreement between two sets of categorical labelings while taking into account agreement occurring by chance. We utilize it as an accuracy metric taking into account the class imbalance present in our data. It is defined as follows:
\begin{align}
    \textit{Cohen's kappa coefficient} = \frac{p_o - p_e}{1 - p_e},
\end{align}
with $p_o$ as the observed agreement and $p_e$ the agreement by chance. For prediction matrix $\mathbf{pred}$ and ground-truth matrix $\mathbf{gt}$, each of size $N$, we compute the observed agreement $p_o$ as the accuracy: ${p_o = \frac{1}{N}\sum_{i=1}^N\mathbf{1}[pred_i = gt_i]}$. We compute the agreement by chance $p_e$ by multiplying the total counts of each state (thus assuming statistical independence) in the ground-truth labels and the predictions: ${p_e = \frac{1}{N^2} \sum_{s \in \mathbf{s}}N_{pred_s} N_{gt_s}}$, where $N_{pred_s}$ and $N_{gt_s}$ denote the prediction and ground-truth counts for state $s$, respectively. Intuitively, we can interpret the metric as the ratio expressing the gain of our classifier relative to random guessing. A value of 1 corresponds to perfect predictions, with 0 corresponding to random guessing (and negative values to worse-than-random performance).

\begin{figure}[t]
\begin{center}\includegraphics[width=1\linewidth]{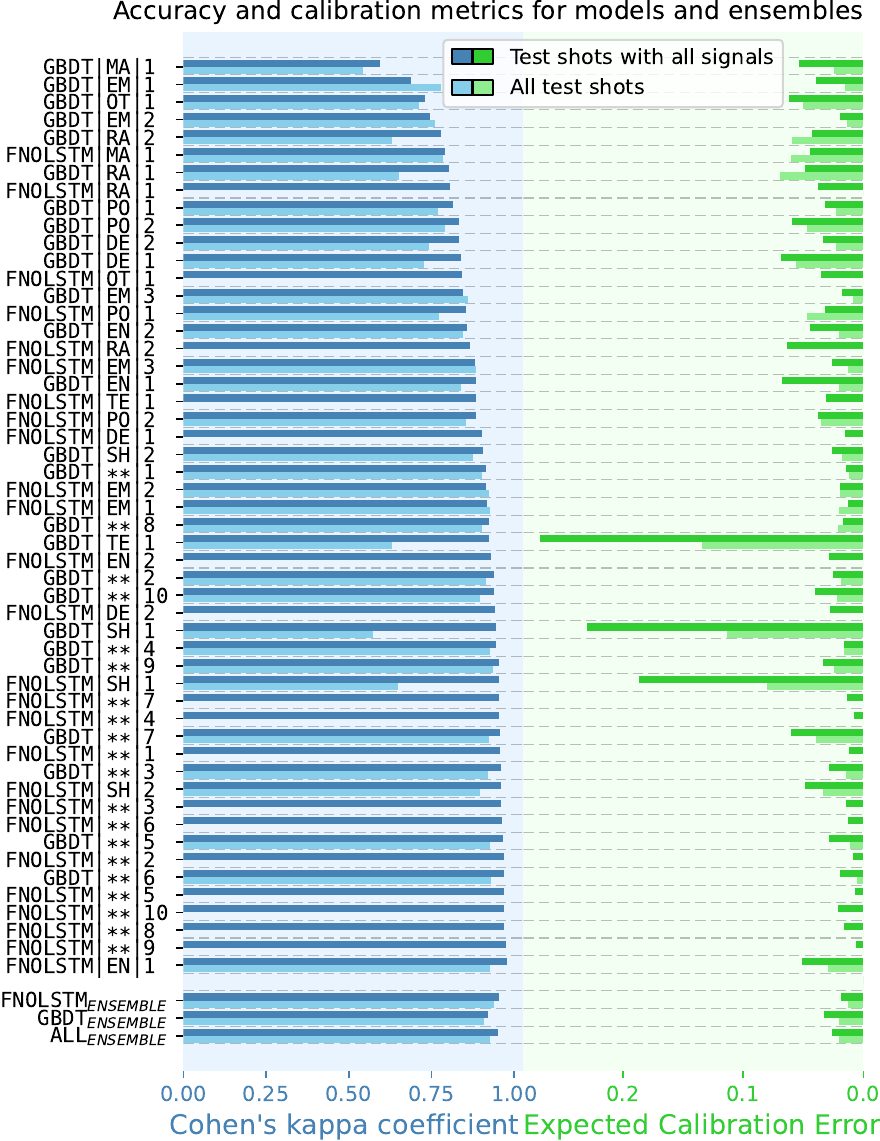}\end{center}
    \caption{An overview of classification performance and uncertainty calibration for all individual models in the ensemble (level 2), ensembles for all FNOLSTM-based and GBDT-based models, and an ensemble of all models (level 1). We plot the Cohen's kappa coefficients on the left (higher is better) and the Expected Calibration Error on the right (lower is better). Feature categories are described by the first two letters, or `$\ast\ast$' for a mix of categories. The different subsets are enumerated for brevity, see~\ref{ap:feature_sets} for details. For each metric, we plot the results for all 34 shots in the test set (bottom bar) and a common subset of 15 shots that contains all features (top bar). The latter set is used to allow a comparison between all individual models: some models cannot provide predictions for all shots in the test set due to missing signals, which is indicated by the lack of a bar in the plot.} %
    \label{fig:all_metrics}%
\end{figure}

\begin{figure}[t]
\begin{center}\includegraphics[width=.825\linewidth]{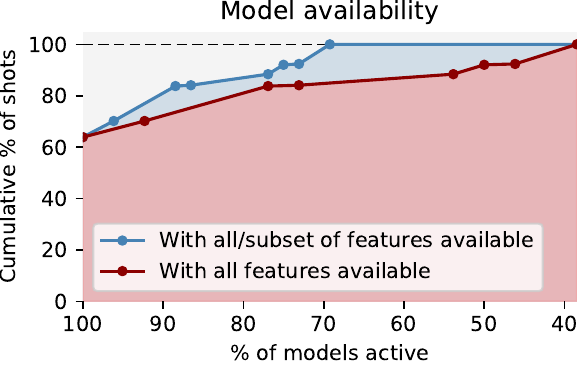}\end{center}
    \caption{The distribution of model availability. That is, we plot the fraction of models that are available, which corresponds to their input features being present in the discharge, for fractions of the dataset. We evaluate the full dataset rather than the test set to give a better sample of signal availability.  The two lines distinguish models that use all signals they are trained on (red) and models that can still run on a discharge, albeit with a reduced input set (blue). The latter correponds to the GBDT models, since they can naturally deal with missing input features. Only approximately $\approx$64\% of discharges have each utilized signal present, resulting in all models being available. In the worst cases, only $\approx$38\% of models are active.}
    \label{fig:availability}%
\end{figure}

The \textit{Expected Calibration Error}~\cite{degroot1983,naeini2015} aims to measure the calibration of a model's confidence outputs. We can express the calibration error as the difference in expectation between the model confidence and accuracy. The ECE approximates this expectation using finite samples by binning the prediction- and confidence-outputs and computing the per-bin confidence/accuracy difference. It is defined as follows: 
\begin{align}
    \hspace{-.11cm}\textit{ECE} = \sum_{m=1}^M \frac{N_m}{N}|\textit{accuracy}(\mathbf{B}_m) - \textit{confidence}(\mathbf{B}_m)|
\end{align}
for $M$ interval bins, together covering the \numrange{0}{1} range of confidences: the $m^\text{th}$ bin covers interval ${(\frac{m-1}{M}, \frac{m}{M}]}$. $\mathbf{B}_m$ denotes the elements in the $m^\text{th}$ bin, $N_m$ the number of elements in $\mathbf{B}_m$, and $N$ the total number of elements. For each bin, $\textit{accuracy}(\mathbf{B}_m)$ is computed as the fraction of correctly predicted samples to total samples, and $\textit{confidence}(\mathbf{B}_m)$ as the average confidence output. Intuitively, the \textit{ECE} can directly be interpreted as an error of the confidence output: an \textit{ECE} of 0.01 indicates that on average, the model's confidence differs from its actual accuracy by 1 percentage point.

\textbf{Overview of results.} An overview of prediction and calibration performance on the test set, measured through Cohen's kappa coefficient and the ECE, is provided in Figure~\ref{fig:all_metrics}; see Table~\ref{tab:allresults} for a tabular version. We plot the results for all \textit{(model + feature set)} settings (level 2) and ensembles for all FNOLSTM models, all GBDT models and for all models (level 1). To be able to compare all feature sets, we provide the metrics on both the full test dataset of 34 discharges and a subset of 15 discharges for which all features are available. The results are sorted on the prediction performance on the latter subset. To give an idea of the  general availability of signals, and subsequently of models using them, we plot the distribution of model availability in Figure~\ref{fig:availability}.

\begin{figure}[t]
\begin{center}\includegraphics[width=.85\linewidth]{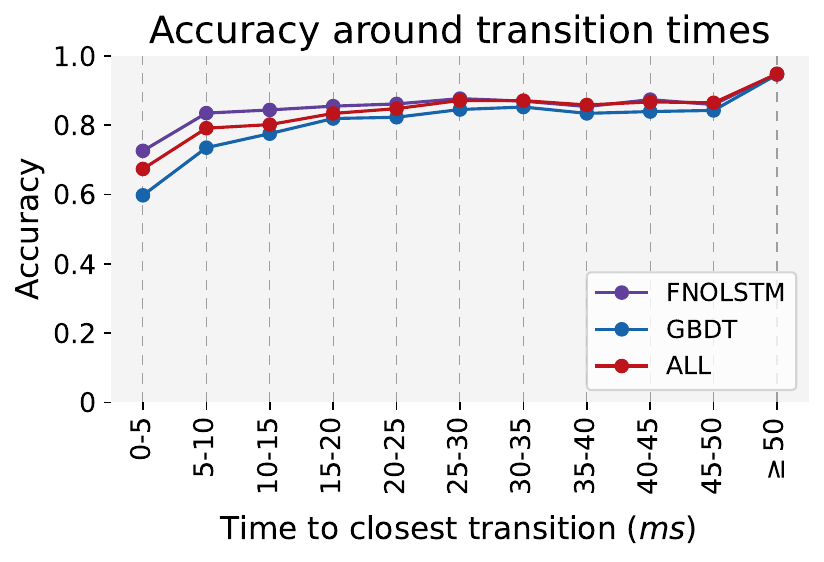}\end{center}
    \caption{The prediction accuracy near the time of state transitions. Each point corresponds to accuracy for timesteps filtered using a \SI{5}{\milli\second} window with varying offsets before and after a transition, with the exception of the last point capturing the remaining timesteps. We observe a significant drop in accuracy in small windows around transitions, with the largest impact for the static-only GBDT ensemble.}
    \label{fig:transition_acc}%
\end{figure}

Perhaps unsurprisingly, mixed feature settings generally provide the best predictive performance and lowest calibration error when they can be applied. FNOLSTM models based on emission and energy content signals also show strong performance while still being applicable to all shots. Notably, mixed feature set-based GBDT models still show strong performance even when applied to all test shots, where some components of the model are disabled due to missing input signals. The ensembled models do not necessarily always beat the \textit{(model + feature set)} models on individual metrics, however they can always be applied and we found them to give more robust predictions and uncertainty estimates.

Lastly, we investigate the relative performance of the FNOLSTM-only, GBDT-only, and complete ensemble with regards to transition times. Specifically, we filter the test set on various incremental windows of~\SI{5}{\milli\second} within~\SI{50}{\milli\second} of state transitions and plot the accuracy on this subset of the data, see Figure~\ref{fig:transition_acc}. Note that we consider accuracy as a metric for easier interpretability\footnote{Subsets of data just before/after transitions naturally have (approximately) balanced ratios of two classes, removing the need to account for class imbalance in the metric.}. While on a large scale the ensembles all perform well, the accuracies drop significantly very close to the transition time, primarily within \numrange{0}{5}\SI{}{\milli\second}; more precise estimates of the exact time of transition are an interesting avenue for future work.

\begin{figure}[t]
\begin{center}\includegraphics[width=1\linewidth]{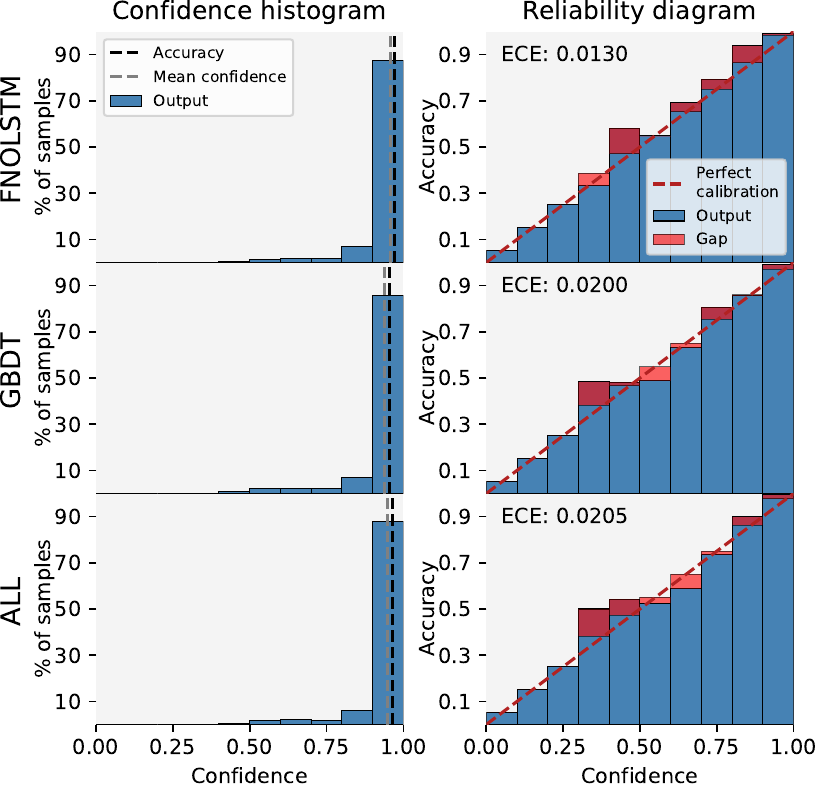}\end{center}
    \caption{The calibration of ensembles of FNOLSTM models, GBDT models and all models. The reliability diagrams (right) show how the model confidence, binned into intervals of 0.1, corresponds to the expected accuracy from the respective bin: a visual respresentation of the ECE. The distribution of model confidences are plotted for context (left).}
    \label{fig:calibration}%
\end{figure}

\begin{figure}[t]
\begin{center}\includegraphics[width=1\linewidth]{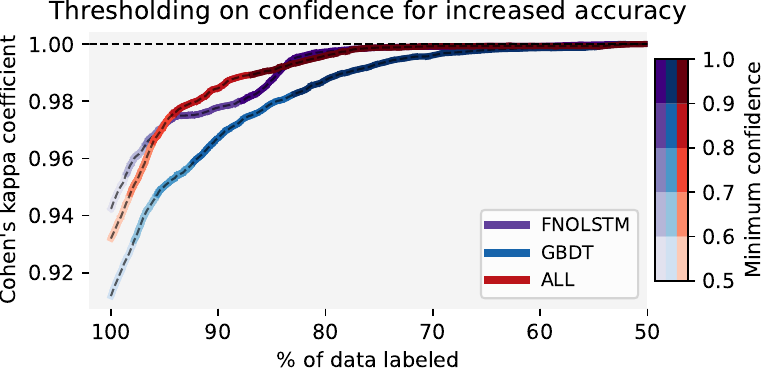}\end{center}
    \caption{The relation between the prediction performance, the model confidence and the fraction of data labeled, plotted for the ensemble of FNOLSTM models, GBDT models and all models. For each ensemble, the line is colored by the minimum confidence level, and displays what fraction of the dataset is still covered, and at what accuracy. By setting this threshold, we can select to label a subset of the dataset with near-perfect results.}
    \label{fig:threshold}%
\end{figure}

\begin{figure*}[t]
\begin{center}\includegraphics[width=.5\linewidth]{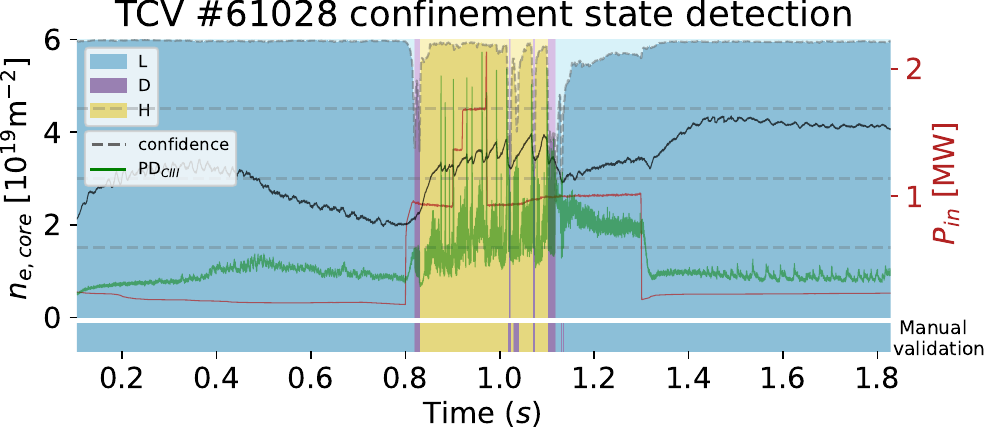}\includegraphics[width=.5\linewidth]{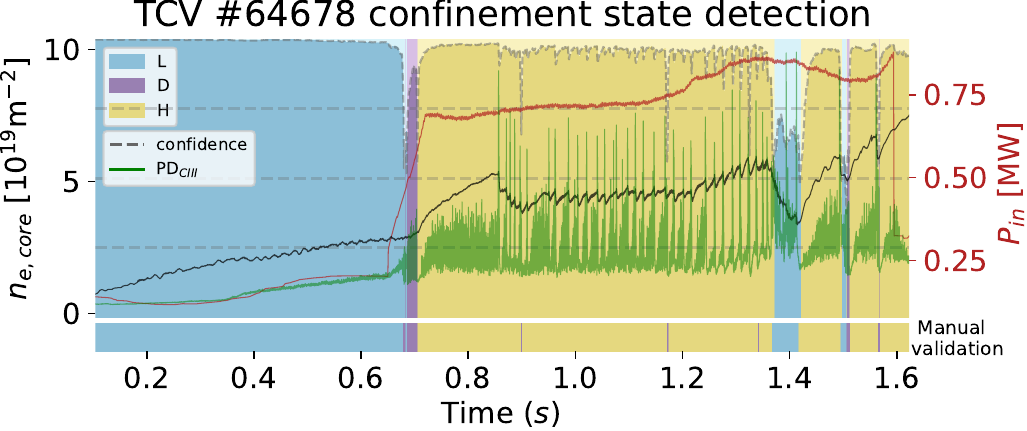}\\\includegraphics[width=.5\linewidth]{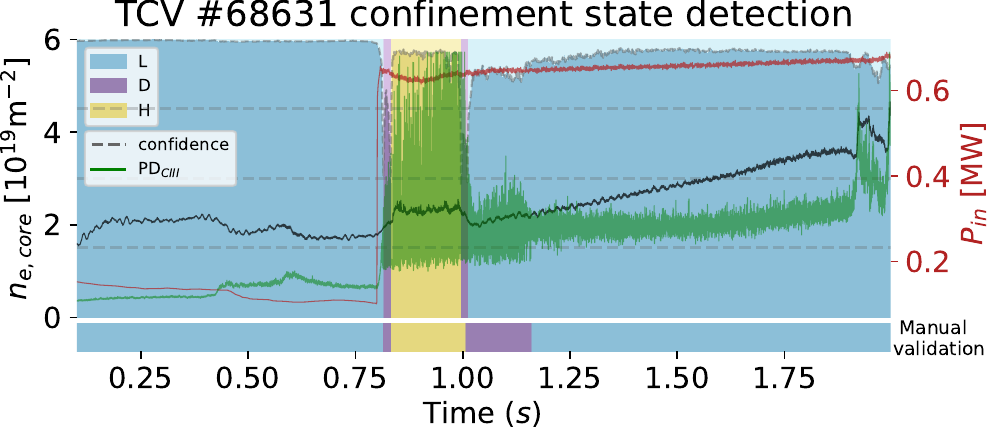}\includegraphics[width=.5\linewidth]{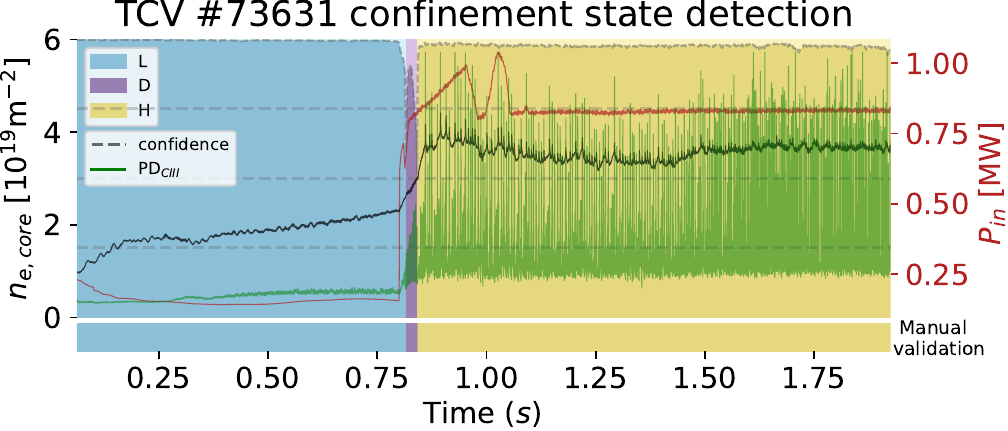}\end{center}
    \caption{Plots illustrating the predictions from the full ensemble. For each discharge we plot the core line integrated density from the interferometer $n_{e,\text{core}}$ (black), the total input power $P_{\textit{in}}$ (red), and overlay the emissions from the photodiode $\text{PD}_{\textit{CIII}}^{}$ (green). The top panel is colored by the ensemble predictions of the confinement state, the bottom stripe indicates the expert's manual labeling for reference. Additionally, the ensemble confidence is provided in the top panel, by the dashed line and the change in background brightness. This scale is always normalized, i.e., the top indicates 1.0 confidence and the bottom 0.0 (confidence quartiles given by dashed horizontal lines).}
    \label{fig:s_test}%
\end{figure*}

\textbf{Confidence-accuracy relationship.} To evaluate the validity of the confidence outputs, we start by plotting reliability diagrams for the FNOLSTM, GBDT and full ensembles in Figure~\ref{fig:calibration}. Generally, all ensembles are relatively well calibrated, with only a few percentage points of calibration error on average. However, since the ensembles are generally very accurate, the majority of confidence predictions fall in a tight range, i.e. $\approx$\numrange{0.93}{0.97}: we cannot trust the ECE in isolation. Fortunately, also around lower confidence estimates the reliability diagrams generally indicate good calibration; more qualitative evaluations are provided in subsequent sections.

Additionally, we explore using the confidence estimates as a threshold for prediction outputs. For example, if one wants to build a database of confinement states but does not necessarily care about fully labeling each discharge, one could filter on high-confidence timeslices to get more reliable results. This relation between prediction accuracy when filtering on a minimum level of confidence, alongside the fraction of test data that remains, is depicted in Figure~\ref{fig:threshold}. We see that the accuracy rises steadily as the threshold is increased, with the full ensemble showing the most advantageous relation. Data can be labeled with little to no errors with $\approx$75\% of timeslices remaining.

\subsection{Qualitative results}\label{sec:qualitative}
\textbf{General overview.} To give an idea of the average performance, we plot 4 discharges from the test set in Figure~\ref{fig:s_test}. These discharges cover various scenarios, e.g. power scans for edge dynamics, high performance scenario development and control near operational limits. In general, there is good agreement between the ensemble and the manual validation, even capturing several transient transitions. Some are missed however, for example in \#61028 and \#64678; nevertheless, there is usually a drop in confidence aligning with the respective transients. The only major error is in \#68631 where a region of dithering is incorrectly labeled as L-mode, albeit at a slightly reduced confidence.

\begin{figure}[t]
\begin{center}\includegraphics[width=1\linewidth]{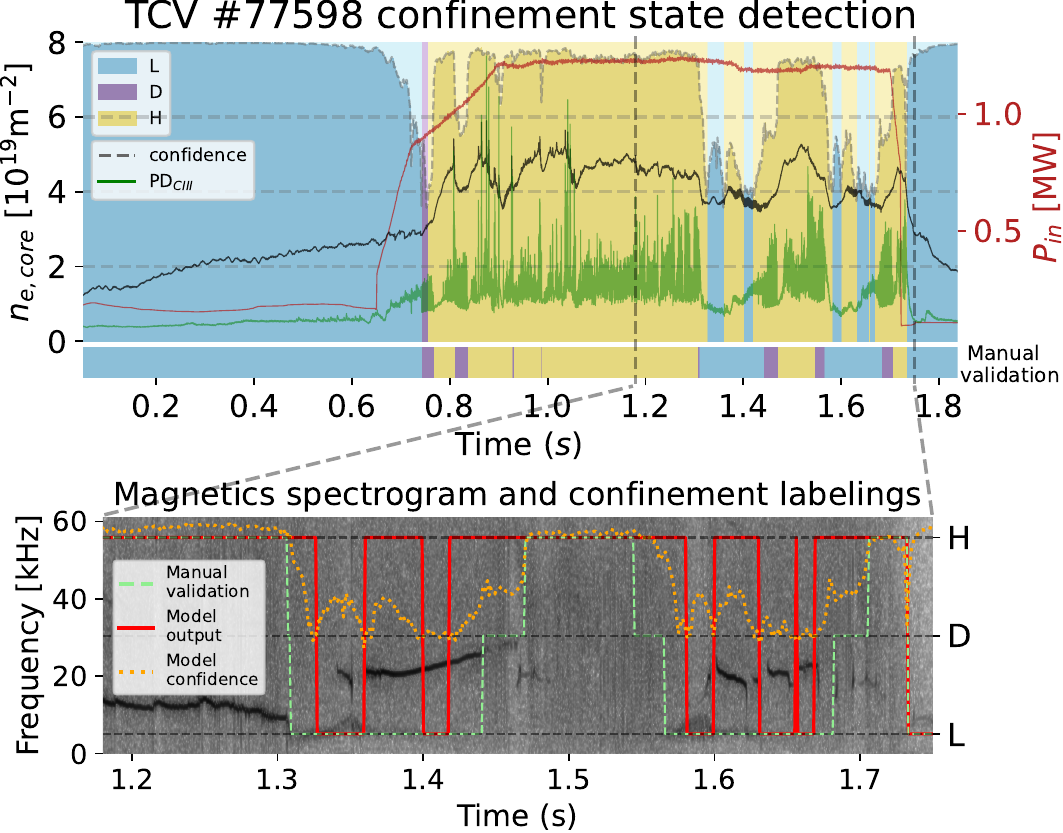}\end{center}
    \caption{The worst result on the test set: \#77598, a discharge testing vertical instability growth rate control. We plot the full discharge, predictions and signals at the top. At the bottom, we zoom in on the region of biggest mismatch, showing the labels, predictions and confidences overlaid on a spectrogram of high frequency magnetic measurements. Multiple incorrectly classified transitions seem to align with MHD activity.}
    \label{fig:s_test_worst}%
\end{figure}

The worst result in the test set is on \#77598, depicted in Figure~\ref{fig:s_test_worst} (top). In this discharge, a radial proximity controller was tested to control the vertical instability growth rate~\cite{marchioni2024}, leading to some unconventional plasma dynamics. Specifically, the controller was active from \SI{1.25}{\second} to \SI{1.70}{\second}, corresponding to the region of biggest mismatch. Nevertheless, an automated labeling method should deal with any scenario. To further investigate, we display the spectrogram from high frequency magnetics for the period of biggest mismatch in Figure~\ref{fig:s_test_worst} (bottom). The prediction errors partially correspond to the occurrence of magnetohydrodynamic (MHD) perturbations, for example around \SI{1.35}{\second} and \SI{1.6}{\second}. Potentially, the effects of these perturbations lead to signature behavior in input signals similar to those in H-mode. Such confusion could be alleviated by also including MHD markers as input signals. Regardless, we see that the confidence also drops low in the regions of mismatch. For example between {$\approx$\numrange{1.4}{1.6}\SI{}{\second}} we have a period where the method predicts H-mode for too many timesteps, but the confidence is high only for the period where it matches the manual validation. Similarly, the H-L back transition at \SI{1.3}{\second} is predicted too late, however the confidence drops where the transition actually occurs. %

\begin{figure}[t]
\begin{center}\includegraphics[width=1\linewidth]{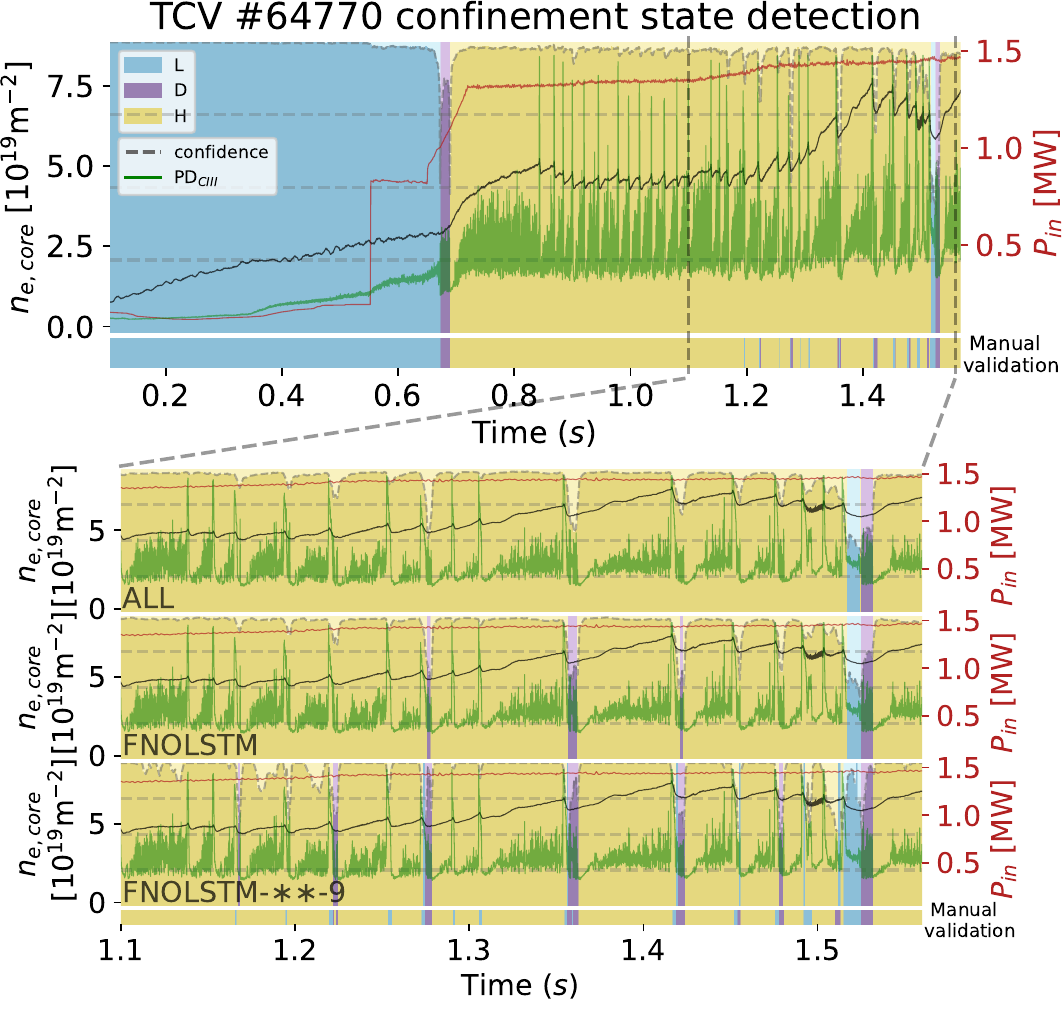}\end{center}
    \caption{Results on \#64770, an IBL scenario development discharge. The full discharge, predictions and signals are plotted at the top. At the bottom, we zoom in on a region with several fast back transitions. Here, we evaluate three sets of predictions: the ensemble of all models, an ensemble of all FNOLSTM models, and the FNOLSTM-$\ast\ast$-9 setting. All models show confidence spikes around the transient events. The dynamic models (bottom two) tend to perform better on these fast transients, albeit with noisier confidence estimates.}
    \label{fig:s_ibl}%
\end{figure}

\textbf{ITER Baseline (IBL) example.} To evaluate a representative scenario we test the ensemble on \#64770, see Figure~\ref{fig:s_ibl}. Specifically, \#64770 is an ITER Baseline scenario development discharge using ECRH power to prevent neoclassical tearing modes~\cite{labit2024}. We see that the main L-D-H phases are matched precisely and with high confidence. Of interest is the phase from \SI{1.1}{\second} onward, where many fast back transitions occurred. To better evaluate the characteristics of the different models, we zoom in for 3 predictions (Figure~\ref{fig:s_ibl} bottom): the full ensemble, the ensemble of all FNOLSTM models, and one of the best-performing \textit{(model + feature set)} settings. All models capture the main H-mode phase and all show dips in the uncertainty corresponding to the transient events. The two models discarding the static formulation perform better around these transients. This discrepancy is not surprising, given that a fast back-and-forth between different states is likely easier to capture by using a context window of signal data that covers the before and after, rather than an individual timeslice only in one state. The individual model (FNOLSTM-$\ast\ast$-9) shows the best performance when it comes to the output labels, but the confidence estimates are a lot noisier, especially from $\SI{1.1}{\second}$ to $\SI{1.25}{\second}$. As such, while individual level 2 \textit{(model + feature set)} ensemble predictions are worth evaluating in scenarios with fast dynamics, these predictions are likely not as robust. Also, they come with a stronger requirement on signal availability--e.g., the FNOLSTM-$\ast\ast$-9 setting could only be applied to 16 out of 34 test discharges due to missing inputs.

\begin{figure}[h]
\begin{center}\includegraphics[width=1\linewidth]{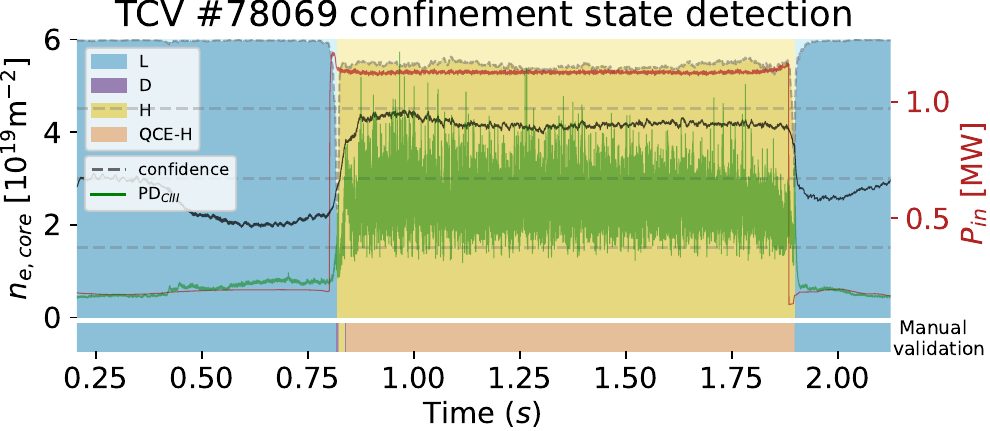}\\\includegraphics[width=1\linewidth]{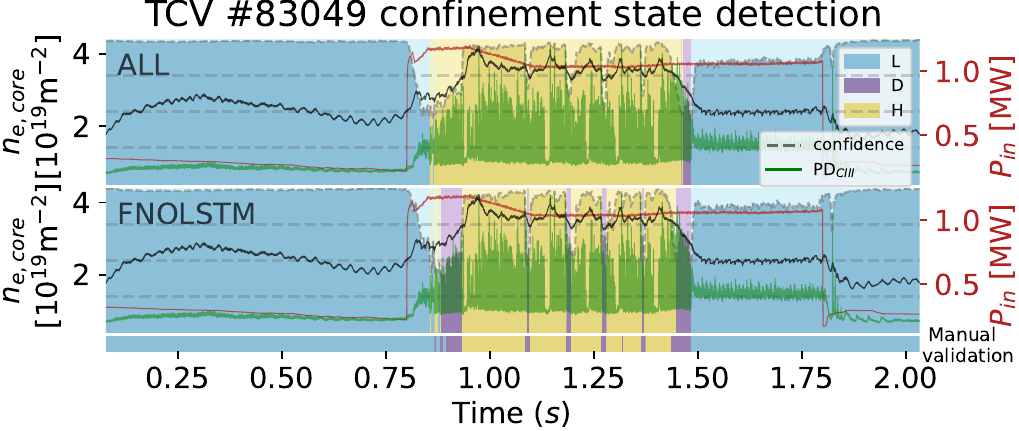}\end{center}
    \caption{Results for two quasi-continuous exhaust scenario development discharges, a successful and a borderline example. For \#78069 we show results for the full ensemble, for \#83049 we show results for both the full ensemble and the FNOLSTM-only ensemble. In the setting of fast transients, the dynamic formulation-only ensembles tend to be more robust.}
    \label{fig:s_qce}%
\end{figure}

\textbf{Alternative scenarios.} Next, we test on discharges from underrepresented scenarios to evaluate challenging circumstances. First, we consider two shots for scenario development of the quasi-continuous exhaust regime~\cite{labit2019}, see Figure~\ref{fig:s_qce}. Shot \#78069 successfully reached the desired small ELM regime, with its time window accurately predicted by the ensemble. In discharge \#83049 it is not as clear, with more ELMy and dithering regions. We plot results for the full ensemble and the FNOLSTM ensemble, illustrating that with more transient behavior using only the dynamic formulation tends to give more precise results. Additionally, we test two negative triangularity discharges~\cite{coda2022}, see Figure~\ref{fig:s_nt} for plots. The ensemble is not sensitive to the non-standard configuration and accurately predicts L-mode for the entire shots' durations. 

\subsection{Extrapolation and robustness}
\label{sec:extrapolation_robustness}
\begin{figure}[h]
\begin{center}\includegraphics[width=1\linewidth]{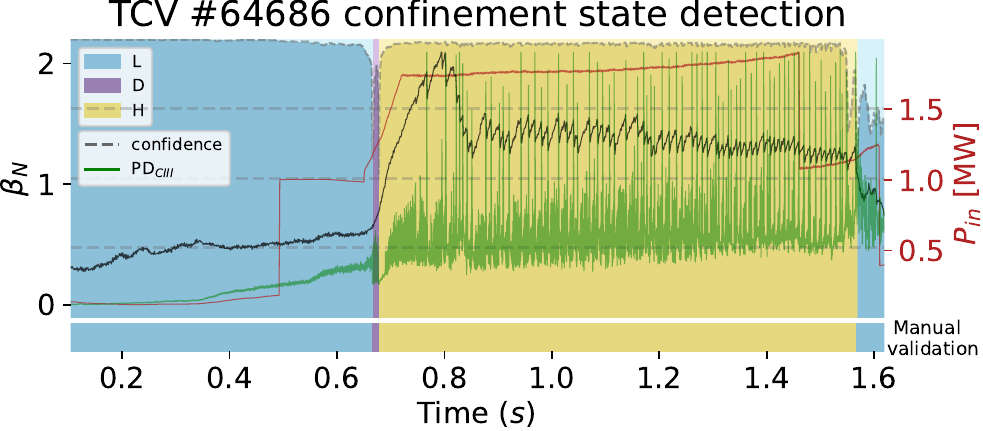}\\\includegraphics[width=1\linewidth]{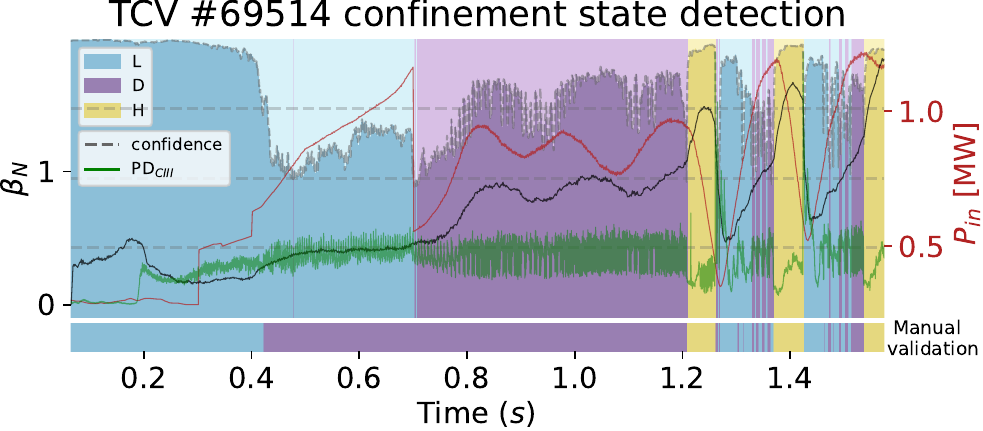}\end{center}
    \caption{Results for two discharges with out-of-distribution conditions. Specifically, we removed discharges with phases with an average $\beta_N > 1.7$ and $\delta_{\text{top}} > 0.3$ for at least \SI{100}{\milli\second} from the `train-validation' set. In general the ensemble still performs well, with the exception of mislabeling the starting time of a long period of dithering.}
    \label{fig:s_betan}%
\end{figure}

\textbf{Out-of-distribution regimes.} To test the ensemble in an out-of-distribution setting, we filtered on shots with an average $\beta_N > 1.7$ and $\delta_{\text{top}} > 0.3$ for a phase of \SI{100}{\milli\second} and removed them from the `train-validation' set, ensuring we do not train on these conditions. Predictions for two shots from this set are given in Figure~\ref{fig:s_betan}. In general, the ensemble still performs well in these conditions, with the exception of mislabeling a dithering region in \#69514 around \SI{0.4}{\second} to \SI{0.7}{\second}, albeit with a low confidence score. In general, the results on this set did not seem significantly different from the remaining test-set discharges. It should be noted that while no discharges with the specified condition were used for training, it is likely that similar discharges were still present in the training data: we are not necessarily evaluating a completely novel scenario.

\begin{figure}[t]
\begin{center}\includegraphics[width=1\linewidth]{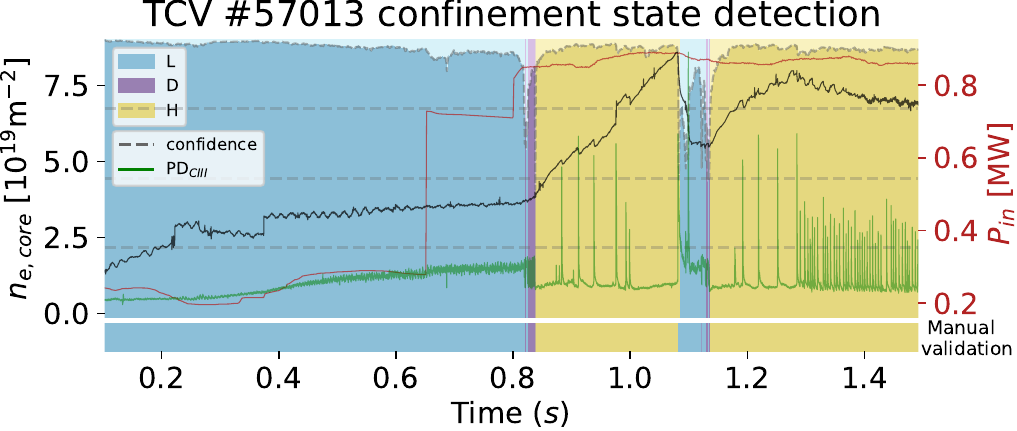}\\\includegraphics[width=1\linewidth]{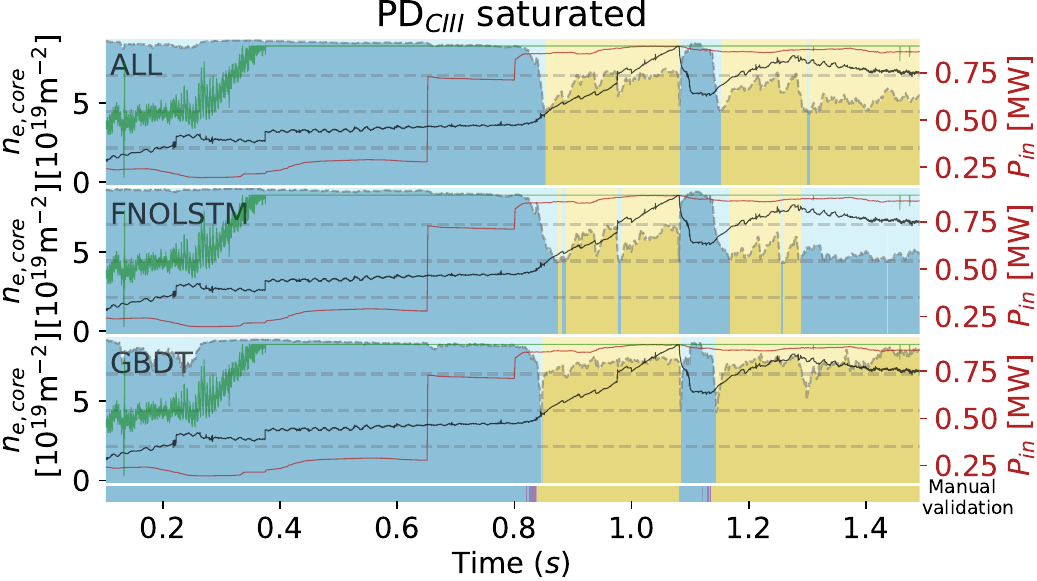}\\\includegraphics[width=1\linewidth]{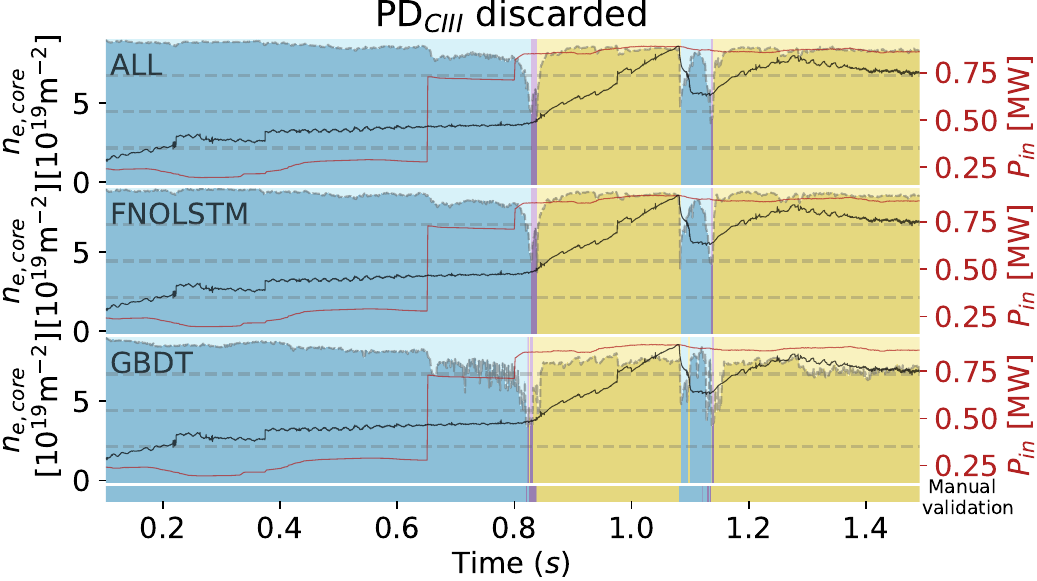}
\end{center}
    \caption{Evaluating the robustness of the ensembles in case of broken or missing signals, specifically for $\text{PD}_{\textit{CIII}}^{}$ measurements. At the top we show the reference discharge \#57013 with the full ensemble predictions. In the middle we show result of a saturated $\text{PD}_{\textit{CIII}}^{}$, a failure mode when inadequate gains are set during operation. Here, the dynamic formulation struggles the most, in contrast to earlier results: including multiple formulations noticeably increases performance. At the bottom we consider the case of removing models using $\text{PD}_{\textit{CIII}}^{}$ from the ensembles: here, the models mostly recover, although with less accuracy compared to the unaffected setting at the top.}
    \label{fig:pdsat}%
\end{figure}

\textbf{Broken or missing signals.} To evaluate the robustness of the ensembling method in more detail, we consider the case of a faulty or missing $\text{PD}_{\textit{CIII}}^{}$ signal. The $\text{PD}_{\textit{CIII}}^{}$ emissions are a key indicator of the confinement state due to its line-of-sight crossing the divertor region: emission patterns such as ELMs leave clear signatures. It is used directly in the dynamic-formulation models, and by both model types through derived spectral features as $\text{PD}_{\text{FFT}}^{}$. The baseline of no errors is given in Figure~\ref{fig:pdsat} (top). Here, the ensemble accurately matches the expert labels. In Figure~\ref{fig:pdsat} (middle) we show the results when $\text{PD}_{\textit{CIII}}^{}$ is saturated, an error mode caused by the diagnostic system's gain being set too large. In contrast to previous examples, the FNOLSTM ensemble now shows the worst results, evidently having a larger dependence on this signal. The GBDT ensemble loses accuracy around the transitions but still provides accurate predictions with high confidence in the stable phases. As a consequence, the full ensemble relies more on the GBDT predictions and still predicts the main phases correctly, highlighting the benefits of the multiple model formulation approach in the full ensemble. Lastly, we consider the case of discarding the signal entirely, i.e., discarding all models utilizing it, in Figure~\ref{fig:pdsat} (bottom). Here, all ensembles mostly recover, although they still lack in precision around the transitions. We note that in this shot there are also small fringe jumps--incorrect interferometer measurements caused by an error in determining the phase difference~\cite{murari2006}--present in the electron core density signal $n_{e,\text{core}}$. They do not seem to significantly affect predictions, although we do see some spikes around the confidence coinciding with the fringe jump times, e.g. \SI{1.0}{\second} for the FNOLSTM in the middle plot. %

\begin{figure}[t]
\begin{center}\includegraphics[width=.9\linewidth]{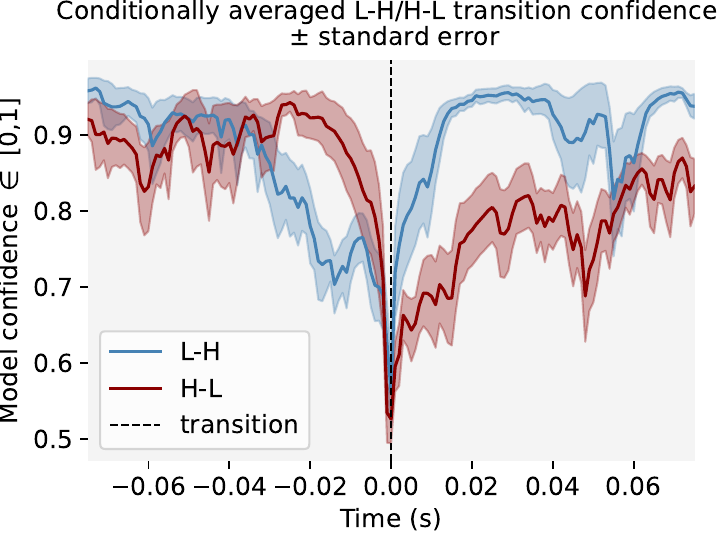}\end{center}
    \caption{Conditionally averaged prediction confidence around L-H and H-L transitions, for a sample of 9 transitions each.}
    \label{fig:lh_hl}%
\end{figure}
\begin{figure}[t]
\begin{center}\includegraphics[width=.9\linewidth]{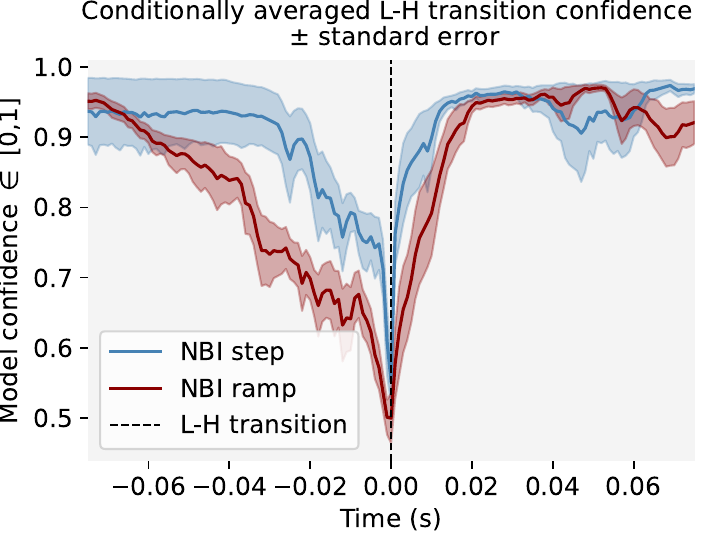}\end{center}
    \caption{Conditionally averaged prediction confidence around L-H transitions with steps or ramps in the NBI, for a sample of 9 transitions each.}
    \label{fig:nbi_ramp}%
\end{figure}

\subsection{Confidence evaluation around state transitions}\label{sec:conditional_average}
For a more statistically robust evaluation of the qualitative behavior of the confidence predictions, we conditionally average the full-ensemble confidences around different types of transitions. Specifically, we compare L-H and H-L transitions, and compare L-H transitions with steps in the NBI power to L-H transitions with more gradual ramps in the input power.

\textbf{L-H vs. H-L transition.} The conditionally averaged model confidences for L-H versus H-L transitions are given in Figure~\ref{fig:lh_hl}. We consider a time window of \SI{150}{\milli\second} around the transition, using L-H and H-L transition pairs from the same shots to minimize other variations in the plasma configuration. There are clear differences in the behavior of the confidence between the two cases. One possible explanation could be that the L-H transition is often more controlled: the model confidences hint at a more gradual phase, and a clear H-mode right after the transition. In contrast, the H-L back transition could be more sudden and uncontrolled, leading to high confidence up to the moment of transition as there are no signs of the transition approaching, along with more uncertainty about the plasma state right after it occurred. Further evaluating these differences in confidence behavior is an interesting avenue for future work.

\textbf{L-H transition with NBI steps vs. NBI ramps.} The conditionally averaged model confidences for steps in the NBI power, compared to ramps in the NBI power, are provided in Figure~\ref{fig:nbi_ramp}. We select 9 shots for each type, with NBI steps denoting a time of at most \SI{20}{\milli\second} between the minimum and maximum power, and ramps a gradual increase for a time window of at least \SI{150}{\milli\second}. The behavior of the confidence reflects the different input power dynamics: the discharges with a step increase show a more sudden drop compared to the ramped increases. There is more model uncertainty around the transition time, with slower, more marginal transitions, consistent with expectations.

\section{Conclusions and Discussion}\label{sec:conclusions}
We have presented a method for the robust classification of confinement states whilst providing meaningful confidence estimates. The method is based on a hierarchical ensemble, combining different types of models and sets of input features on the top level. On the second level, these model/feature combinations consist of a mini-ensemble trained on different data-splits, which are averaged and empirically calibrated such that we can meaningfully combine them on the top level.

We have evaluated the approach quantitatively and qualitatively on a variety of scenarios. In most cases, the full ensemble provided accurate predictions and meaningful uncertainty estimates. However, especially in transient cases, ensembles of FNOLSTM-only models tended to perform slightly better. This advantage came at the cost of more sensitivity to corrupted input features. 

The approach gives the flexibility of choosing different components given the desired use-case. For example, if one wants to robustly identify main phases for large datasets, the full ensemble is more suitable given its added robustness. If one is specifically interested in back-and-forth transitions the FNOLSTM-ensemble is more suitable. Additionally, if one is aware of errors in certain diagnostics, one could disable the subset of models using this signal as an input feature.

The main weakness of the method lies in estimating the precise time of transition. While it robustly identifies main phases of various discharges and generally finds all the main transitions, the accuracy drops in small time windows ($<$\SI{5}{\milli\second}) around transitions or around fast transients. Future efforts focusing specifically on the transition time, rather than a general purpose classifier, are of interest to address this weakness.

\subsection{Future Work}
A potential avenue to increasing performance around transition regions would be to reformulate the problem to detecting the time of a transition, akin to change point detection methods~\cite{aminikhanghahi2016}, rather than labeling every timestep. This type of model could be used in a cascaded setting with the current approach: first, we robustly detect the main phases for the different confinement states. Then, a specialized model refines the prediction around the time of transition. Additionally, one could explore a wider set of neural network architectures to improve performance, e.g. for the local pattern extractor~\cite{ho2020,Liu2021ICCV,kovachki2021neural} or for the long term correlations~\cite{vaswani2017,beck2024xlstm,gu2024mamba}.

Another interesting avenue for future research is multi-device confinement state classification. Especially in light of future devices, one will not have access to a large set of discharges in order to create a dataset. Additionally, even if the experiments are available, accurate labeling of the confinement states is a time-consuming effort. Initial efforts in this direction have been made~\cite{marceca2021}, however, at a significant drop in performance: the fundamental differences in timescales and dynamics prove a significant challenge. Potential approaches to tackle these issues consider transfer learning~\cite{zhuang2021transfer}, physics-based normalization for input signals or device-invariant model architectures. Notably, there has been significant progress on cross-machine data-driven models for disruption prediction, for example when training on one device and evaluating on another~\cite{katesharbeck2019} or by adding only a limited number of shots from a new device~\cite{zhu021disr,zheng2023disr}.

A real-time version of the proposed method is of interest for control applications. In principle only minor adaptations would have to be made. The FNOLSTM architecture is structurally similar to a prior NN-based confinement state classifier~\cite{matoslhd2020}, which is already implemented in the TCV control system~\cite{marceca2021,galperti2024}. The computational cost of the tree ensemble method is also real-time compatible. All individual ensemble components can run in parallel, with the final ensembling procedure requiring negligible computation time: there should be no latency bottlenecks. The prediction lag parameter in the current models is at most \SI{10}{\milli\second}, although one could choose to lower this parameter at the cost of some precision. Input-wise, one would have to restrict the input set to real-time available signals. Finally, to ensure parity between training and real-time use, one should take care to use causal interpolation methods rather than the linear interpolation used in this paper.

Another interesting direction would be to reformulate the output to a figure of merit for confinement performance, rather than a discrete state label. For example in the real-time setting, one could directly optimize this quantity with model-based control techniques. In a similar vein, one could extend the approach to differentiate different types of H-mode, e.g. distinguishing ELMy and ELM-free regimes; see~\cite{zorek2022,gill2024} for related works towards this setting.

More generally, the ensembling strategy could be applied to problems of similar structure, such as disruption prediction~\cite{katesharbeck2019}. The joint integration of robustness to signal issues and uncertainty quantification makes it a good candidate for real-time prediction strategies, where reliability and interpretability are crucial for integration in disruption avoidance control schemes~\cite{galperti2024}.

\section*{Acknowledgements}
The authors would like to thank Christian Donner and Giulio Romanelli for insightful discussions. This work was funded in part by a Swiss Data Science Center project grant (C21-14). This work has been carried out within the framework of the EUROfusion Consortium, partially funded by the European Union via the Euratom Research and Training Programme (Grant Agreement No 101052200 — EUROfusion). The Swiss contribution to this work has been funded in part by the Swiss State Secretariat for Education, Research and Innovation (SERI). Views and opinions expressed are however those of the author(s) only and do not necessarily reflect those of the European Union, the European Commission or SERI. Neither the European Union nor the European Commission nor SERI can be held responsible for them. This work was supported in part by the Swiss National Science Foundation. This work used the Dutch national e-infrastructure with the support of the SURF Cooperative using grant no. EINF-7709.

\section*{References}
\bibliographystyle{plainurl_abrev}
\bibliography{main}

\clearpage
\appendix
\renewcommand{\thesection}{\appendixname~\Alph{section}}
\onecolumn
\setcounter{footnote}{1}
\counterwithin{figure}{section}
\counterwithin{lstlisting}{section}
\renewcommand{\thefigure}{\Alph{section}.\arabic{figure}}
\renewcommand{\thetable}{\Alph{section}.\arabic{table}}
\section{Feature sets}\label{ap:feature_sets}
\begin{figure}[h]
\begin{center}\includegraphics[width=0.5\linewidth]{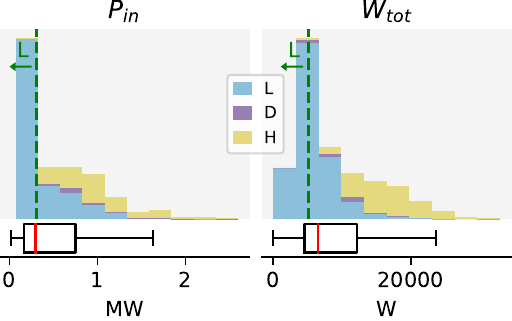}\end{center}
    \caption{Distributions of the total input power and the plasma stored energy in the dataset, following the same procedure as Figure~\ref{fig:dataset_eda}. We overlay the `all L-mode' thresholds (L$^{\textit{fraction}}_{0.99}$) for the two features: below this threshold value, at least 99\% of the timeslices are in L-mode.}
    \label{fig:thresholds_example}%
\end{figure}

The feature sets in the ensembling procedure are selected by the feature categorization and their individual discriminative power. To quantify the latter, we fit small models on each individual feature. Specifically, we fit a depth-2 decision tree to classify L, D or H-mode timeslice-by-timeslice. The performance is evaluated using Cohen's kappa coefficient~\cite{cohen1960}. Additionally, to identify parameter ranges consistently associated with a specific confinement state, we optimize thresholds on each individual feature that correspond to the largest amount of the data we can classify as `all L-mode' or `all H-mode' with at least 99\% accuracy. In other words, we check whether a feature can individually identify one of the two main confinement states in subparts of the parameter space. For example, total input power $P_{\textit{in}}$ can be used to trivially label some timeslices as L-mode given a minimum power requirement for any H-mode, see also Figure~\ref{fig:thresholds_example} for an illustration. We express this metric as the fraction of the data which can be labeled with such a threshold while keeping at least 99\% accuracy. Additionally, we report the signal availability over all timeslices in the dataset.

The results of all these metrics are provided in Table~\ref{tab:features}, for all features introduced in Table~\ref{tab:signals}. Note that we denote all spectral features computed from the photodiode signals ($\text{PD}_{\text{FFT}}^{}$ in Table~\ref{tab:signals}). The subscript integer denotes the window size in \SI{}{\milli\second} for the sliding window FFT, whereas the postfix $\in\{\textit{p}, \textit{c}, \textit{f}\}$ denotes whether the window is in the \textit{p}ast, \textit{c}entered or in the \textit{f}uture w.r.t. the given timeslice.

The feature sets cover both individual categories and combinations thereof. For each category we construct a model covering either all features or the most discriminative features following their Cohen's kappa coefficient values. The mixed feature sets cover both top-$k$ subsets for each category and rank-$k$ subsets: we both fit models taking in as much information as possible, while also fitting models with no mutual dependencies but still using informative features. Similarly, we select subsets using the threshold-orderings, although fewer in total because a substantial number of features cannot be used for any meaningful thresholding\footnote{Naturally, these features are still useful once combined with other more directly significant features.}, resulting in a value of 0 in Table~\ref{tab:features}. The resulting \textit{(model + feature set)} configurations are given in Table~\ref{tab:model_featuresets}.

The feature sets are identical between the FNOLSTM and GBDT models, with the exception of the photodiode related features. For the FNOLSTM we artificially rank $\text{PD}_{\textit{CIII}}^{}$ and $\text{PD}_{\textit{H}\alpha}^{}$ as the most informative emissions feature. These signals are not absolutely calibrated, making their raw value uninformative for classifying L, D or H-mode. However, the emission patterns they pick up clearly correspond to confinement state-related dynamics such as Edge Localized Modes (ELMs) or dithering cycles. The FNOLSTM-based models can fit these patterns because of their dynamic nature, making it a key feature to include. In contrast, the GBDT-models only take static information, making the raw signal value uninformative; rather, it relies on the constructed spectral features for the photodiode signal. To avoid redundancy in these $\text{PD}_{\text{FFT}}$ features, only the centered-window feature is used for each time window size; the past and future windows are only used in a specific category with all FFT features (FNOLSTM-EM-3 and GBDT-EM-3). 

\newpage

\arrayrulecolor{black}
\def\arrvline{\hfil\kern\arraycolsep\vline\kern-\arraycolsep\hfilneg}
\setlength\tabcolsep{5.3pt}
\begingroup
\centering
\begin{longtable}{lccccp{0.45cm}lcccc}
 & Cohen's & & & Fraction & & & Cohen's & & & Fraction\\[-3pt]
Feature & kappa & L$^{\textit{fraction}}_{0.99}$ & H$^{\textit{fraction}}_{0.99}$ & \makebox[0pt]{available} & & Feature & kappa & L$^{\textit{fraction}}_{0.99}$ & H$^{\textit{fraction}}_{0.99}$ & \makebox[0pt]{available} \\ \cmidrule[\heavyrulewidth]{1-5}\cmidrule[\heavyrulewidth]{7-11}
\addlinespace[-\belowrulesep]
\cellcolor[RGB]{255,210,204} $A_p$ & \cellcolor[RGB]{255.0,255.0,255.0} 0.000 & \cellcolor[RGB]{250.0,253.0,249.6} 0.033 & \cellcolor[RGB]{255.0,255.0,255.0} 0.00000 & \cellcolor[RGB]{179.9,224.4,173.9} 0.997 &  & \cellcolor[RGB]{209,235,255} $\text{PD}_{\text{FFT}_{50\textit{-c}}}^{{\textit{H}\alpha}}$ & \cellcolor[RGB]{201.5,233.2,197.2} 0.704 & \cellcolor[RGB]{254.7,254.9,254.7} 0.002 & \cellcolor[RGB]{244.4,250.7,243.5} 0.00021 & \cellcolor[RGB]{182.9,225.6,177.2} 0.985 \\
\cellcolor[RGB]{255,210,204} $\delta_{\text{bottom}}$ & \cellcolor[RGB]{249.4,252.7,249.0} 0.073 & \cellcolor[RGB]{243.5,250.3,242.6} 0.076 & \cellcolor[RGB]{255.0,255.0,255.0} 0.00000 & \cellcolor[RGB]{179.9,224.4,173.9} 0.997 &  & \cellcolor[RGB]{209,235,255} $\text{PD}_{\text{FFT}_{50\textit{-f}}}^{{\textit{H}\alpha}}$ & \cellcolor[RGB]{202.0,233.4,197.9} 0.697 & \cellcolor[RGB]{254.9,255.0,254.9} 0.001 & \cellcolor[RGB]{255.0,255.0,255.0} 0.00000 & \cellcolor[RGB]{187.0,227.3,181.7} 0.968 \\
\cellcolor[RGB]{255,210,204} $\delta_{\text{top}}$ & \cellcolor[RGB]{255.0,255.0,255.0} 0.000 & \cellcolor[RGB]{244.9,250.9,244.1} 0.066 & \cellcolor[RGB]{255.0,255.0,255.0} 0.00000 & \cellcolor[RGB]{179.9,224.4,173.9} 0.997 &  & \cellcolor[RGB]{209,235,255} $\text{PD}_{\text{FFT}_{100\textit{-p}}}^{{\textit{H}\alpha}}$ & \cellcolor[RGB]{212.2,237.5,208.8} 0.563 & \cellcolor[RGB]{255.0,255.0,255.0} 0.000 & \cellcolor[RGB]{254.7,254.9,254.7} 0.00001 & \cellcolor[RGB]{179.0,224.0,173.0} 1.000 \\
\end{longtable}
\endgroup
\setlength\tabcolsep{6pt}
\arrayrulecolor{black}

\arrayrulecolor[rgb]{0.9,0.9,0.9}
\setlength\tabcolsep{5.3pt}
\begingroup
\centering
\begin{longtable}{l|p{2.7cm}|p{10.9cm}}
Model & Description & Feature set \\\cmidrule[\heavyrulewidth]{1-3}
\addlinespace[-\belowrulesep]
FNOLSTM-SH-1 & Shaping, top 4. & \colorbox[RGB]{255,210,204}{\strut $A_p$}, \colorbox[RGB]{255,210,204}{\strut $\delta_{\text{bottom}}$}, \colorbox[RGB]{255,210,204}{\strut $\kappa$}, \colorbox[RGB]{255,210,204}{\strut $Z_{\text{axis}}$} \\\hline
FNOLSTM-SH-2 & Shaping, all. & \colorbox[RGB]{255,210,204}{\strut $A_p$}, \colorbox[RGB]{255,210,204}{\strut $\delta_{\text{bottom}}$}, \colorbox[RGB]{255,210,204}{\strut $\delta_{\text{top}}$}, \colorbox[RGB]{255,210,204}{\strut $\Delta_{\text{in}}$}, \colorbox[RGB]{255,210,204}{\strut $\Delta_{\text{out}}$}, \colorbox[RGB]{255,210,204}{\strut $\kappa$}, \colorbox[RGB]{255,210,204}{\strut $R_0$}, \colorbox[RGB]{255,210,204}{\strut $a$}, \colorbox[RGB]{255,210,204}{\strut $R_{\text{axis}}$}, \colorbox[RGB]{255,210,204}{\strut $Z_{\text{axis}}$}, \colorbox[RGB]{255,210,204}{\strut $V_p$} \\\hline
FNOLSTM-EM-1 & Emission, top 4. & \colorbox[RGB]{209,235,255}{\strut $\text{PD}_{\textit{CIII}}^{}$}, \colorbox[RGB]{209,235,255}{\strut $\text{PD}_{\textit{H}\alpha}^{}$}, \colorbox[RGB]{209,235,255}{\strut $\text{PD}_{\text{FFT}_{20\textit{-c}}}^{{\textit{H}\alpha}}$}, \colorbox[RGB]{209,235,255}{\strut $\text{PD}_{\text{FFT}_{50\textit{-c}}}^{{\textit{H}\alpha}}$} \\
\end{longtable}
\endgroup
\setlength\tabcolsep{6pt}
\arrayrulecolor{black}

\newpage

\clearpage\section{Model training}\label{ap:model_training}
\subsection{FNOLSTM}
The training procedure of FNOLSTM models is summarized as follows. At timestep $m$, for time window $w$, stride size $p$ and prediction offset $k$, the neural network maps the input window $\mathbf{x}_{t_{m-w}:t_m}$, combined with previous hidden state $(\mathbf{h}_{t_{m-p}}, \mathbf{c}_{t_{m-p}})$, to predictions $\mathbf{\widetilde{y}}^{\mathbf{s},t_{m-k}}$. We use categorical cross-entropy as the loss function between the predictions and ground truth values: 
\begin{align}
\mathcal{L}(\mathbf{y}^{\mathbf{s},t_{m-k}}, \mathbf{\widetilde{y}}^{\mathbf{s},t_{m-k}}) = -\sum_{s \in \mathbf{s}} y^{s,t_{m-k}} \log \hat{y}^{s,t_{m-k}},\label{eq:cce}
\end{align}
equivalent to minimizing the negative log-likelihood. We use standard gradient-based optimization as provided by PyTorch~\cite{paszke2019}; specifically, we use the schedule-free Adam optimizer~\cite{defazio2024}. 

One parameter update step consists of computing the loss between a mini-batch of signal timeseries and the corresponding confinement state labels. We iteratively predict confinement states as we slide over the input signal with stride size $p$ while continuously updating the hidden state $(\mathbf{h}_{t_{m-p}}, \mathbf{c}_{t_{m-p}})$. The number of these unrolling steps is a hyperparameter, see also~\ref{ap:parameters}. The loss computation and parameter update is done for the entire batch + unrolling at once, that is, applying backpropagation through time~\cite{werbos1988}.

We sample each element in the batch as a random \textit{(shot + initial timestep)} combination. To account for the imbalance in the different class labels, we precompute the total number of L, D and H labels for all \textit{(shot + initial timestep)} combinations given the chosen stride size, prediction offset, and number of unrolling steps. Then, rather than uniformly sampling the \textit{(shot + initial timestep)} inputs, we reweight the sampling probability based on the total probabilities of the L, D, and H labels across all elements to ensure that all confinement states are sampled equally often in expectation. We optimize for $\approx$100 batches per epoch for 50 epochs. After every 10 epochs we compute Cohen's kappa coefficient on the validation discharges and save the model with the minimum validation error.

\subsection{GBDT}
The training procedure of GBDT models is summarized as follows. At timestep $m$ we map input signals $\mathbf{x}^{\mathbf{u},t_m}$ to predictions $\mathbf{\widetilde{y}}^{\mathbf{s},t_m}$. As loss function we again use the categorical cross-entropy between the predictions and the reference labels $\mathbf{y}^{\mathbf{s},t_m}$, see Equation~\ref{eq:cce}. The random forest is build through gradient boosting using XGBoost~\cite{xgboost2016}. Trees are fit one at a time utilizing the full train set for each tree, with new trees minimizing the residual error w.r.t. the already-trained trees.

The samples cover \textit{(shot + timestep)} combinations in the training dataset. Specifically, we subsample timesteps with L or H labels to account for the class imbalance. We utilize various subsampling ratios, see~\ref{ap:parameters} for details. Additionally, regardless of the aforementioned subsampling, we always sample all 50 timesteps before and after each confinement state transition, to ensure these more difficult timeslices are always part of the dataset.

\clearpage\section{Settings and hyperparameters}\label{ap:parameters}
\subsection{FNOLSTM}
The NN model architecture consists of an FNO-based encoder, an LSTM unit and a small MLP for the final prediction, see also Equation~\ref{eq:fdynamic_fnolstm}. We use a fixed structure of layers and optimize the layer hyperparameters. Certain layer sizes are coupled to reduce the search space and to avoid extremely unbalanced models; for this, we introduce parameters FNO$_\text{feat}$, FNO$_\text{mode}$ and LSTM$_\text{feat}$. The architecture is as follows. The input consists of a time window of size $w$ for $\textit{Nu}$ input features, i.e. $\mathbf{x} \in \mathbb{R}^{w \times \textit{Nu}}$. This $\mathbf{x}$ is transformed by an FNO layer to FNO$_\text{feat}$ hidden features utilizing FNO$_\text{mode}$ modes and a ReLU\footnote{$\sigma(z) = \text{max}(0, z).$} nonlinearity. We follow with another FNO layer, mapping to $2 \cdot \text{FNO}_\text{feat}$ hidden features using FNO$_\text{mode}$ modes and a ReLU nonlinearity. We apply a dropout~\cite{dropout2014} of 0.5 and do max-pooling of 2 over the temporal axis. Then, we flatten the temporal axis and map with a linear layer to LSTM$_\text{feat}$ features followed by ReLU; this output of the temporal feature extractor is denoted as $z$. We transform $z$ using an LSTM\footnote{Also taking as input the previous LSTM's hidden state.} with a hidden size of LSTM$_\text{feat}$. The result is summed to $z$, i.e. the residual connection, and mapped to $\frac{1}{3}\text{LSTM}_\text{feat}$ features with a ReLU nonlinearity. We apply dropout, map to 3 features and apply softmax\footnote{$\sigma(z)_i = \frac{e^{z_i}}{\sum_{j=1}^{n} e^{z_j}}$.} function, the output prediction $\mathbf{y} \in \mathbb{R}^3$.

Non-architecture parameters we optimize are the time window size $w$ and the prediction offset $k$ (see Equation~\ref{eq:fdynamic_fnolstm}). Stride parameter $p$ is fixed to 10, which results in a prediction rate of \SI{1}{\kilo\hertz} given that the signals are (re)sampled at \SI{10}{\kilo\hertz}. All neural networks are optimized with schedule-free Adam~\cite{defazio2024} following the training procedure describe in~\ref{ap:model_training}; we optimize the learning rate and the number of warmup steps for the optimizer, and optimize the batch size\footnote{We keep the number of batches per epoch fixed despite the larger batch size to keep the total number of update steps fixed.} and the number of unrolling steps done for each sample during training.

\subsection{GBDT}
The GBDT models utilize the implementation provided by XGBoost~\cite{xgboost2016}. For each model we optimize the number of decision trees in the forest, the maximum depth of a single decision tree, and the learning rate of the gradient boosting algorithm. Additionally, we optimize the ratios of resampling the different classes. We always subsample L and H-mode timeslices to match the number of D timeslices; here, we optimize whether we do this rebalancing at a 1:1, 3:1 or 5:1 ratio for L and H w.r.t. D. Note that we always sample all timeslices around transition points regardless of class rebalancing, see also~\ref{ap:model_training}.

\subsection{Optimization procedure and parameter ranges}
All hyperparameters are optimized with Bayesian optimization through Optuna~\cite{optuna2019}, with the Cohen's kappa coefficient~\cite{cohen1960} on validation-set discharges as the target metric. They are shared on the level of the 4 folds used for each \textit{(model + feature set)} combination. Consequently, each hyperparameter trial considers training 4 models, with the resulting validation metrics averaged out. For each \textit{(model + feature set)} configuration we do 100 trials for FNOLSTM-based models, and 250 trials for GBDT-based models. The hyperparameter ranges for both model types are provided in Tables~\ref{tab:hyperparamnn} and~\ref{tab:hyperparamxgboost}, respectively. 

\begin{table}[h]
\centering
\begin{tabular}{p{4cm}p{4cm}}
Hyperparameter & Range (type) \\
\cmidrule[\heavyrulewidth]{1-2}
Time window $w$ & \numrange{20}{100} (\textit{int}, step: 10) \\
Prediction offset $k$ & 20--$w$ (\textit{int}, step: 10) \\
Learning rate & \numrange{0.0005}{0.2} (\textit{float}, log-scale) \\
Warmup steps & \numrange{0}{150} (\textit{int}) \\
Batch size & \numrange{16}{384} (\textit{int}) \\
Train unrolling steps & \numrange{10}{80} (\textit{int}, step: 5) \\
FNO$_\text{feat}$ & \numrange{14}{68} (\textit{int}, step: 6) \\
FNO$_\text{mode}$  & \numrange{8}{14} (\textit{int}, step: 2) \\
LSTM$_\text{feat}$ & \numrange{14}{68} (\textit{int}, step: 6) \\
\end{tabular}
\caption{Hyperparameter ranges for all FNOLSTM-based models. }
\label{tab:hyperparamnn}
\end{table}

\begin{table}[h]
\centering
\begin{tabular}{p{4cm}p{4cm}}
Hyperparameter & Range (type) \\
\cmidrule[\heavyrulewidth]{1-2}
Learning rate & \numrange{0.01}{1.0} (\textit{float}, log-scale) \\
No. trees & \numrange{100}{1500} (\textit{int}, step: 20) \\
Max. tree depth & \numrange{3}{12} (\textit{int}) \\
L/H to D ratio & $\{1, 3, 5\}$ (\textit{int}) \\
\end{tabular}
\caption{Hyperparameter ranges for all GBDT-based models.}
\label{tab:hyperparamxgboost}
\end{table}

\clearpage\section{Extra results}\label{ap:results}
\begin{figure}[h]
\begin{center}\includegraphics[width=0.5\linewidth]{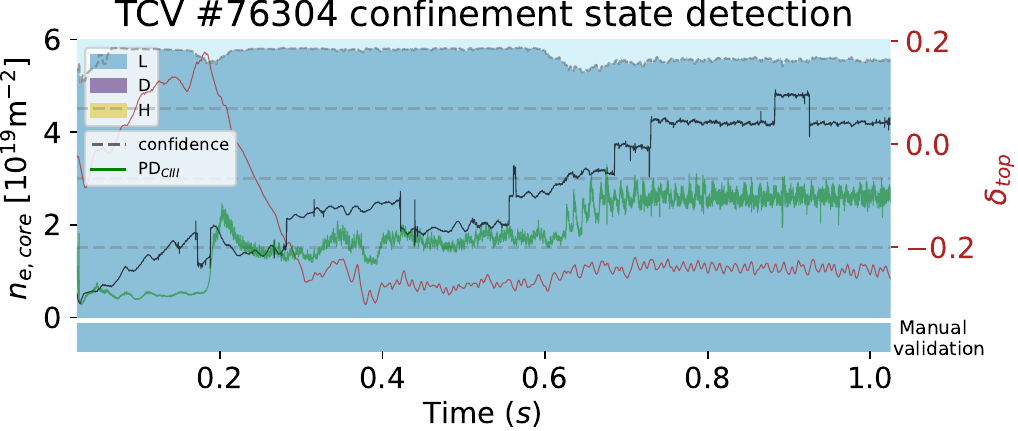}\includegraphics[width=0.5\linewidth]{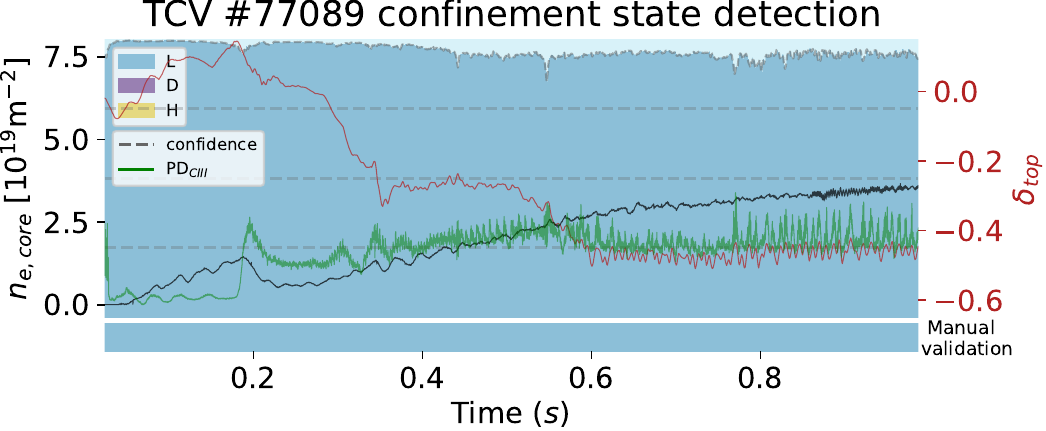}\end{center}
    \caption{Two examples of prediction results in a negative triangularity scenario. The ensemble accurately predicts L-mode with high confidence, even in a non-standard magnetic configuration not commonly present in the dataset. We also note the presence of fringe jumps~\cite{murari2006} in the interferometer signal in \#76304: the ensembling approach effectively mitigates sensitivity to the corrupted signal.}
    \label{fig:s_nt}%
\end{figure}

This appendix contains extra figures or tables for the model evaluation. Specifically, Table~\ref{tab:allresults} contains metric values for Cohen's kappa score~\cite{cohen1960} and the Expected Calibration Error~\cite{degroot1983,naeini2015} for all individual \textit{(model + feature set)} configurations and the FNOLSTM-only, GBDT-only and full ensemble, similar to Figure~\ref{fig:all_metrics}. Figure~\ref{fig:s_nt} depicts full-ensemble predictions on two negative triangularity discharges from the test set.

\arrayrulecolor{black}
\def\arrvline{\hfil\kern\arraycolsep\vline\kern-\arraycolsep\hfilneg}
\setlength\tabcolsep{5.3pt}
\begingroup
\centering
\begin{longtable}{lccccc}

 & \multicolumn{2}{c}{Cohen's kappa coefficient}  & \multicolumn{2}{c}{Expected Calibration Error} \\\cmidrule[\heavyrulewidth]{2-6}
Model & All test (34) & Subset (15) & All test (34) & Subset (15)\\
\cmidrule[\heavyrulewidth]{1-6}
\addlinespace[-\belowrulesep]
FNOLSTM-SH-1 & \cellcolor[RGB]{245.8,251.2,245.1} 0.648 & \cellcolor[RGB]{187.8,227.6,182.4} 0.954 & \cellcolor[RGB]{253.9,222.3,211.6} 0.0798 & \cellcolor[RGB]{252.2,172.2,145.2} 0.1867\\
FNOLSTM-SH-2 & \cellcolor[RGB]{198.7,232.1,194.3} 0.896 & \cellcolor[RGB]{186.8,227.2,181.4} 0.959 & \cellcolor[RGB]{254.6,244.0,240.4} 0.0336 & \cellcolor[RGB]{254.4,236.8,230.9} 0.0488\\
FNOLSTM-EM-1 & \cellcolor[RGB]{193.0,229.7,188.1} 0.926 & \cellcolor[RGB]{194.8,230.5,190.1} 0.917 & \cellcolor[RGB]{254.8,250.3,248.7} 0.0201 & \cellcolor[RGB]{255.0,253.7,253.2} 0.0128\\
FNOLSTM-EM-2 & \cellcolor[RGB]{193.7,230.0,188.9} 0.923 & \cellcolor[RGB]{195.1,230.6,190.4} 0.915 & \cellcolor[RGB]{254.9,250.6,249.1} 0.0195 & \cellcolor[RGB]{254.8,250.4,248.9} 0.0198\\
FNOLSTM-EM-3 & \cellcolor[RGB]{200.9,232.9,196.6} 0.885 & \cellcolor[RGB]{201.4,233.1,197.1} 0.882 & \cellcolor[RGB]{255.0,253.9,253.5} 0.0124 & \cellcolor[RGB]{254.7,247.3,244.8} 0.0265\\
FNOLSTM-MA-1 & \cellcolor[RGB]{220.2,240.8,217.4} 0.783 & \cellcolor[RGB]{218.6,240.2,215.8} 0.791 & \cellcolor[RGB]{254.2,231.7,224.0} 0.0598 & \cellcolor[RGB]{254.5,238.9,233.7} 0.0443\\
FNOLSTM-DE-1 & \cellcolor[RGB]{255, 255, 255} - & \cellcolor[RGB]{197.5,231.5,192.9} 0.903 & \cellcolor[RGB]{255, 255, 255} - & \cellcolor[RGB]{254.9,252.3,251.5} 0.0157\\
FNOLSTM-DE-2 & \cellcolor[RGB]{255, 255, 255} - & \cellcolor[RGB]{190.3,228.6,185.2} 0.941 & \cellcolor[RGB]{255, 255, 255} - & \cellcolor[RGB]{254.7,246.7,244.0} 0.0277\\
FNOLSTM-TE-1 & \cellcolor[RGB]{255, 255, 255} - & \cellcolor[RGB]{201.0,233.0,196.7} 0.884 & \cellcolor[RGB]{255, 255, 255} - & \cellcolor[RGB]{254.7,245.1,241.9} 0.0311\\
FNOLSTM-PO-1 & \cellcolor[RGB]{222.0,241.5,219.4} 0.774 & \cellcolor[RGB]{206.7,235.3,202.9} 0.854 & \cellcolor[RGB]{254.4,237.8,232.2} 0.0467 & \cellcolor[RGB]{254.7,244.6,241.2} 0.0322\\
FNOLSTM-PO-2 & \cellcolor[RGB]{206.5,235.2,202.7} 0.855 & \cellcolor[RGB]{200.9,232.9,196.6} 0.885 & \cellcolor[RGB]{254.6,243.3,239.5} 0.0349 & \cellcolor[RGB]{254.6,242.1,237.9} 0.0376\\
FNOLSTM-EN-1 & \cellcolor[RGB]{193.1,229.7,188.2} 0.926 & \cellcolor[RGB]{183.1,225.7,177.4} 0.978 & \cellcolor[RGB]{254.7,245.8,242.8} 0.0297 & \cellcolor[RGB]{254.4,235.8,229.6} 0.0509\\
FNOLSTM-EN-2 & \cellcolor[RGB]{255, 255, 255} - & \cellcolor[RGB]{192.5,229.5,187.6} 0.929 & \cellcolor[RGB]{255, 255, 255} - & \cellcolor[RGB]{254.7,246.5,243.7} 0.0282\\
FNOLSTM-RA-1 & \cellcolor[RGB]{255, 255, 255} - & \cellcolor[RGB]{215.7,239.0,212.6} 0.807 & \cellcolor[RGB]{255, 255, 255} - & \cellcolor[RGB]{254.6,242.1,237.8} 0.0376\\
FNOLSTM-RA-2 & \cellcolor[RGB]{255, 255, 255} - & \cellcolor[RGB]{204.6,234.5,200.7} 0.865 & \cellcolor[RGB]{255, 255, 255} - & \cellcolor[RGB]{254.2,230.1,221.9} 0.0632\\
FNOLSTM-OT-1 & \cellcolor[RGB]{255, 255, 255} - & \cellcolor[RGB]{209.3,236.3,205.7} 0.841 & \cellcolor[RGB]{255, 255, 255} - & \cellcolor[RGB]{254.6,243.0,239.1} 0.0357\\
FNOLSTM-$\ast\ast$-1 & \cellcolor[RGB]{255, 255, 255} - & \cellcolor[RGB]{187.1,227.3,181.7} 0.958 & \cellcolor[RGB]{255, 255, 255} - & \cellcolor[RGB]{255.0,253.9,253.6} 0.0123\\
FNOLSTM-$\ast\ast$-2 & \cellcolor[RGB]{255, 255, 255} - & \cellcolor[RGB]{185.2,226.5,179.7} 0.967 & \cellcolor[RGB]{255, 255, 255} - & \cellcolor[RGB]{255.0,255.0,255.0} 0.0090\\
FNOLSTM-$\ast\ast$-3 & \cellcolor[RGB]{255, 255, 255} - & \cellcolor[RGB]{186.6,227.1,181.2} 0.960 & \cellcolor[RGB]{255, 255, 255} - & \cellcolor[RGB]{254.9,252.9,252.3} 0.0144\\
FNOLSTM-$\ast\ast$-4 & \cellcolor[RGB]{255, 255, 255} - & \cellcolor[RGB]{187.5,227.5,182.2} 0.955 & \cellcolor[RGB]{255, 255, 255} - & \cellcolor[RGB]{255.0,255.0,255.0} 0.0079\\
FNOLSTM-$\ast\ast$-5 & \cellcolor[RGB]{255, 255, 255} - & \cellcolor[RGB]{185.0,226.5,179.5} 0.968 & \cellcolor[RGB]{255, 255, 255} - & \cellcolor[RGB]{255.0,255.0,255.0} 0.0070\\
FNOLSTM-$\ast\ast$-6 & \cellcolor[RGB]{255, 255, 255} - & \cellcolor[RGB]{186.3,227.0,180.9} 0.961 & \cellcolor[RGB]{255, 255, 255} - & \cellcolor[RGB]{255.0,253.6,253.1} 0.0131\\
FNOLSTM-$\ast\ast$-7 & \cellcolor[RGB]{255, 255, 255} - & \cellcolor[RGB]{187.6,227.5,182.3} 0.955 & \cellcolor[RGB]{255, 255, 255} - & \cellcolor[RGB]{254.9,253.4,252.9} 0.0134\\
FNOLSTM-$\ast\ast$-8 & \cellcolor[RGB]{255, 255, 255} - & \cellcolor[RGB]{184.9,226.4,179.4} 0.969 & \cellcolor[RGB]{255, 255, 255} - & \cellcolor[RGB]{254.9,252.3,251.4} 0.0158\\
FNOLSTM-$\ast\ast$-9 & \cellcolor[RGB]{255, 255, 255} - & \cellcolor[RGB]{183.8,226.0,178.2} 0.975 & \cellcolor[RGB]{255, 255, 255} - & \cellcolor[RGB]{255.0,255.0,255.0} 0.0063\\
FNOLSTM-$\ast\ast$-10 & \cellcolor[RGB]{255, 255, 255} - & \cellcolor[RGB]{184.9,226.4,179.4} 0.969 & \cellcolor[RGB]{255, 255, 255} - & \cellcolor[RGB]{254.8,249.7,247.9} 0.0214\\\hline
GBDT-SH-1 & \cellcolor[RGB]{255.0,255.0,255.0} 0.574 & \cellcolor[RGB]{189.5,228.3,184.4} 0.945 & \cellcolor[RGB]{253.4,206.6,190.8} 0.1133 & \cellcolor[RGB]{252.0,166.0,137.0} 0.2303\\
GBDT-SH-2 & \cellcolor[RGB]{202.7,233.7,198.6} 0.875 & \cellcolor[RGB]{196.8,231.3,192.2} 0.906 & \cellcolor[RGB]{254.9,251.3,250.1} 0.0179 & \cellcolor[RGB]{254.7,247.3,244.8} 0.0264\\
GBDT-EM-1 & \cellcolor[RGB]{221.3,241.3,218.7} 0.777 & \cellcolor[RGB]{238.3,248.2,236.9} 0.688 & \cellcolor[RGB]{254.9,252.7,251.9} 0.0149 & \cellcolor[RGB]{254.5,241.0,236.5} 0.0398\\
GBDT-EM-2 & \cellcolor[RGB]{224.4,242.5,222.0} 0.761 & \cellcolor[RGB]{227.3,243.7,225.1} 0.746 & \cellcolor[RGB]{254.9,253.3,252.7} 0.0137 & \cellcolor[RGB]{254.9,250.6,249.2} 0.0193\\
GBDT-EM-3 & \cellcolor[RGB]{205.8,234.9,201.9} 0.859 & \cellcolor[RGB]{208.6,236.1,204.9} 0.844 & \cellcolor[RGB]{255.0,255.0,255.0} 0.0085 & \cellcolor[RGB]{254.9,251.5,250.3} 0.0176\\
GBDT-MA-1 & \cellcolor[RGB]{255.0,255.0,255.0} 0.544 & \cellcolor[RGB]{255.0,255.0,255.0} 0.595 & \cellcolor[RGB]{254.8,248.1,245.9} 0.0247 & \cellcolor[RGB]{254.3,234.5,227.9} 0.0537\\
GBDT-DE-1 & \cellcolor[RGB]{231.1,245.3,229.2} 0.726 & \cellcolor[RGB]{209.8,236.6,206.2} 0.838 & \cellcolor[RGB]{254.3,233.3,226.3} 0.0563 & \cellcolor[RGB]{254.1,227.6,218.6} 0.0686\\
GBDT-DE-2 & \cellcolor[RGB]{228.2,244.1,226.1} 0.741 & \cellcolor[RGB]{210.8,237.0,207.3} 0.833 & \cellcolor[RGB]{254.8,249.0,247.0} 0.0228 & \cellcolor[RGB]{254.6,244.0,240.5} 0.0334\\
GBDT-TE-1 & \cellcolor[RGB]{249.1,252.6,248.6} 0.631 & \cellcolor[RGB]{193.4,229.9,188.5} 0.924 & \cellcolor[RGB]{253.0,196.6,177.6} 0.1346 & \cellcolor[RGB]{252.0,166.0,137.0} 0.2694\\
GBDT-PO-1 & \cellcolor[RGB]{222.8,241.9,220.2} 0.770 & \cellcolor[RGB]{214.3,238.4,211.1} 0.814 & \cellcolor[RGB]{254.8,248.9,246.9} 0.0230 & \cellcolor[RGB]{254.6,244.6,241.2} 0.0322\\
GBDT-PO-2 & \cellcolor[RGB]{218.8,240.2,215.9} 0.791 & \cellcolor[RGB]{210.9,237.0,207.4} 0.832 & \cellcolor[RGB]{254.4,237.9,232.3} 0.0466 & \cellcolor[RGB]{254.2,231.9,224.3} 0.0594\\
GBDT-EN-1 & \cellcolor[RGB]{209.4,236.4,205.8} 0.840 & \cellcolor[RGB]{201.2,233.1,196.9} 0.883 & \cellcolor[RGB]{254.8,250.1,248.5} 0.0205 & \cellcolor[RGB]{254.1,227.9,219.1} 0.0678\\
GBDT-EN-2 & \cellcolor[RGB]{208.4,236.0,204.7} 0.845 & \cellcolor[RGB]{206.3,235.1,202.5} 0.856 & \cellcolor[RGB]{254.8,250.1,248.5} 0.0204 & \cellcolor[RGB]{254.5,239.0,233.7} 0.0443\\
GBDT-RA-1 & \cellcolor[RGB]{245.4,251.1,244.7} 0.650 & \cellcolor[RGB]{216.5,239.3,213.4} 0.803 & \cellcolor[RGB]{254.1,227.4,218.4} 0.0690 & \cellcolor[RGB]{254.4,237.0,231.2} 0.0484\\
GBDT-RA-2 & \cellcolor[RGB]{249.3,252.7,248.8} 0.630 & \cellcolor[RGB]{221.0,241.1,218.3} 0.779 & \cellcolor[RGB]{254.2,232.0,224.5} 0.0591 & \cellcolor[RGB]{254.5,239.6,234.6} 0.0429\\
GBDT-OT-1 & \cellcolor[RGB]{233.9,246.4,232.3} 0.711 & \cellcolor[RGB]{230.2,244.9,228.3} 0.730 & \cellcolor[RGB]{254.4,236.2,230.1} 0.0501 & \cellcolor[RGB]{254.2,230.7,222.7} 0.0620\\
GBDT-$\ast\ast$-1 & \cellcolor[RGB]{197.3,231.5,192.8} 0.904 & \cellcolor[RGB]{195.2,230.6,190.5} 0.915 & \cellcolor[RGB]{255.0,254.3,254.0} 0.0116 & \cellcolor[RGB]{254.9,252.9,252.2} 0.0146\\
GBDT-$\ast\ast$-2 & \cellcolor[RGB]{195.4,230.7,190.7} 0.914 & \cellcolor[RGB]{190.8,228.8,185.7} 0.938 & \cellcolor[RGB]{254.9,250.8,249.4} 0.0190 & \cellcolor[RGB]{254.8,247.9,245.6} 0.0252\\
GBDT-$\ast\ast$-3 & \cellcolor[RGB]{194.3,230.2,189.5} 0.919 & \cellcolor[RGB]{186.9,227.2,181.5} 0.958 & \cellcolor[RGB]{254.9,253.1,252.5} 0.0141 & \cellcolor[RGB]{254.7,246.4,243.6} 0.0283\\
GBDT-$\ast\ast$-4 & \cellcolor[RGB]{193.0,229.7,188.1} 0.926 & \cellcolor[RGB]{189.3,228.2,184.1} 0.946 & \cellcolor[RGB]{254.9,252.1,251.1} 0.0163 & \cellcolor[RGB]{254.9,252.3,251.4} 0.0158\\
GBDT-$\ast\ast$-5 & \cellcolor[RGB]{193.2,229.8,188.3} 0.925 & \cellcolor[RGB]{185.3,226.6,179.8} 0.967 & \cellcolor[RGB]{255.0,254.5,254.4} 0.0110 & \cellcolor[RGB]{254.7,246.2,243.4} 0.0287\\
GBDT-$\ast\ast$-6 & \cellcolor[RGB]{192.7,229.6,187.7} 0.928 & \cellcolor[RGB]{185.1,226.5,179.6} 0.968 & \cellcolor[RGB]{255.0,255.0,255.0} 0.0053 & \cellcolor[RGB]{254.9,250.7,249.4} 0.0191\\
GBDT-$\ast\ast$-7 & \cellcolor[RGB]{193.2,229.8,188.4} 0.925 & \cellcolor[RGB]{187.1,227.3,181.8} 0.957 & \cellcolor[RGB]{254.5,241.2,236.8} 0.0394 & \cellcolor[RGB]{254.2,231.7,224.1} 0.0598\\
GBDT-$\ast\ast$-8 & \cellcolor[RGB]{197.7,231.6,193.2} 0.901 & \cellcolor[RGB]{193.4,229.9,188.6} 0.924 & \cellcolor[RGB]{254.8,249.7,247.9} 0.0214 & \cellcolor[RGB]{254.9,251.8,250.8} 0.0168\\
GBDT-$\ast\ast$-9 & \cellcolor[RGB]{191.1,229.0,186.1} 0.936 & \cellcolor[RGB]{187.8,227.6,182.5} 0.954 & \cellcolor[RGB]{254.8,248.3,246.1} 0.0243 & \cellcolor[RGB]{254.6,243.8,240.1} 0.0339\\
GBDT-$\ast\ast$-10 & \cellcolor[RGB]{198.8,232.1,194.4} 0.896 & \cellcolor[RGB]{190.6,228.7,185.5} 0.939 & \cellcolor[RGB]{254.8,249.3,247.4} 0.0222 & \cellcolor[RGB]{254.5,241.0,236.4} 0.0399\\\hline
FNOLSTM$_{\text{ENSEMBLE}}$ & \cellcolor[RGB]{190.7,228.8,185.7} 0.938 & \cellcolor[RGB]{187.6,227.5,182.3} 0.955 & \cellcolor[RGB]{255.0,253.6,253.2} 0.0130 & \cellcolor[RGB]{254.9,250.8,249.4} 0.0189\\
GBDT$_{\text{ENSEMBLE}}$ & \cellcolor[RGB]{196.7,231.2,192.1} 0.907 & \cellcolor[RGB]{194.1,230.2,189.3} 0.921 & \cellcolor[RGB]{254.8,250.3,248.8} 0.0200 & \cellcolor[RGB]{254.6,244.4,240.9} 0.0327\\
ALL$_{\text{ENSEMBLE}}$ & \cellcolor[RGB]{192.7,229.6,187.8} 0.928 & \cellcolor[RGB]{188.1,227.7,182.9} 0.952 & \cellcolor[RGB]{254.8,250.1,248.5} 0.0205 & \cellcolor[RGB]{254.7,247.5,245.1} 0.0260\\

\\
\caption{Classification performance (Cohen's kappa coefficient, higher is better) and uncertainty calibration (Expected Calibration Error, lower is better) for all configurations and ensembles on the test set, for both the full test set (34 shots) and a common subset of test shots that contain all features (15 shots).}\label{tab:allresults}
\end{longtable}
\endgroup
\setlength\tabcolsep{6pt}
\arrayrulecolor{black}

\end{document}